\documentclass[10pt,preprint]{aastex}

\usepackage{amsmath}
\usepackage{amssymb}

\newcommand{\A}[1]{\boldsymbol{#1}}
\newcommand{\sgn}{\textrm{sgn}}
\newcommand{\pd}[2]{\frac{\partial #1}{\partial #2}}

\newcommand{\HALF}{\frac{1}{2}}
\newcommand{\vol}{\mathcal{V}}
\newcommand{\nv}{\hat{\A{n}}}
\newcommand{\DS}{\displaystyle}

\newcommand{\Riemann}{\A{\mathcal{R}}}

\slugcomment{Submitted to the \apj}
\shorttitle{Relativistic Piecewise Parabolic Method}
\shortauthors{Mignone et~al.}
\received{2004 October 19}

\begin{document}

\title{THE PIECEWISE PARABOLIC METHOD FOR \\
       MULTIDIMENSIONAL RELATIVISTIC FLUID DYNAMICS}

\author{
A. Mignone\altaffilmark{1,2,3},
T. Plewa\altaffilmark{2,3,4}, and
G. Bodo\altaffilmark{1}
}

\altaffiltext{1}{INAF Osservatorio Astronomico di Torino, Strada dell'Osservatorio 20, 10025 Pino Torinese, Italy}
\altaffiltext{2}{Department of Astronomy \& Astrophysics,
   The University of Chicago,
   Chicago, IL 60637}
\altaffiltext{3}{Center for Astrophysical Thermonuclear Flashes,
   The University of Chicago,
   Chicago, IL 60637}
\altaffiltext{4}{Nicolaus Copernicus Astronomical Center,
   Bartycka 18,
   00716 Warsaw, Poland}

\begin{abstract}

We present an extension of the Piecewise Parabolic Method to special
relativistic fluid dynamics in multidimensions. The scheme is
conservative, dimensionally unsplit, and suitable for a general
equation of state. Temporal evolution is second-order accurate and
employs characteristic projection operators; spatial interpolation is
piece-wise parabolic making the scheme third-order accurate in smooth
regions of the flow away from discontinuities. The algorithm is
written for a general system of orthogonal curvilinear coordinates and
can be used for computations in non-cartesian geometries.  A
non-linear iterative Riemann solver based
on the two-shock approximation is used in flux calculation. 
In this approximation, an initial
discontinuity decays into a set of discontinuous waves only implying
that, in particular, rarefaction waves are treated as flow
discontinuities. We also present a new and simple equation of state
which approximates the exact result for the relativistic perfect gas
with high accuracy.  The strength of the new method is demonstrated in
a series of numerical tests and more complex simulations in one, two and
three dimensions.

\end{abstract}



\section{INTRODUCTION}
%
%
%
%
%

Highly energetic astrophysical phenomena are known to be, in many
cases, relativistic in nature. A wide range of objects, in fact,
exhibits a number of properties that can be accounted for only in the
framework of the theory of special or general relativity: 
superluminal motion of relativistic jets in extragalactic radio
sources \citep{Begelman+84}, jets and accretion flows around massive
compact objects \citep{Koide+99,Meier+01}, pulsar winds
\citep{DelZanna+04,Bogovalov+05}, gamma ray bursts
\citep{Aloy00,ZWmF03,Mizuno+04}, as well as particle beams produced in
heavy-ion collisions in terrestrial experiments
\citep{Ackermann+01,Morita+02,Molnar+04} characterized by flow
velocities very close ($\gtrsim 0.99 c$) to the speed of light. Such
problems are difficult to study numerically due to inherent complexity
of the problem itself exhibiting rapid spatial and temporal changes
often demanding development of more efficient algorithms.

Recent progress in the development of powerful methods in
computational fluid dynamics allowed researchers to gain understanding
of relativistic flows. Numerical formulations of the special
relativistic fluid equations have been investigated by several authors
in the last decade \citep[for an excellent review
see][]{MM03}. However, there now exists strong evidence that a
particular class of numerical methods, the so-called high-resolution
shock-capturing schemes \citep[HRSC henceforth, e.g.,][]{DFIM98,
MMIMD92}, provide the necessary means to develop stable and robust
relativistic (special or general) fluid dynamics codes.  HRSC methods
rely on the conservative formulation of the fluid equations. This
formulation is of fundamental importance in representing the evolution
of flows with steep gradients and discontinuities. One of the key
aspects of these schemes is the temporal evolution of the system,
which routinely involves (exact or approximate) solution of Riemann
problems at the interfaces separating numerical grid cells, thus
making these schemes highly successful in modeling discontinuous
flows. Besides, these algorithms have at least 2nd order accuracy in
smooth regions of the flow and have been shown to produce excellent
results for a wide variety of problems.

Shock-capturing schemes have a long standing tradition in the
framework of solving the Euler equations of gas dynamics
\citep[e.g.,][]{Colella85,CW84,VanLeer97,Toro97}.  Several of these
``classical'' schemes have now been extended to special relativistic
hydrodynamics \citep[see][and references
therein]{Balsara94,DW97,FK96,DFIM98,SZS01}. The Piecewise Parabolic
Method (PPM) by \citet[CW84 henceforth]{CW84} is still considered as
the state-of-the-art method in computational fluid dynamics.  The
original PPM algorithm has been recently re-formulated for the
relativistic fluid flows in one spatial dimension by \cite{MM96}. Here
we present an extension of the PPM method to multidimensional
relativistic hydrodynamics. As we will show, the main differences
between the classical and relativistic version of PPM are the coupling
between normal and transverse velocity components introduced by the
Lorentz factor and the coupling of the latter to the specific
enthalpy.

Our relativistic PPM consists of several components. The
piece-wise parabolic interpolation is done in the volume coordinate
(CW84) and provides third-order accuracy in space in smooth parts of
the flow. Temporal evolution uses characteristic information and is
second-order accurate. The scheme also uses a non-linear iterative
Riemann solver based on the two-shock approximation. The unsplit
fully-coupled corner-transport upwind method is used for advection
\citep{Colella90,Saltzman94}. Several verifications tests are
presented to demonstrate the correctness of the implementation.

The paper is organized as follows. In \S\ref{sec:equations} we
give a short review of the special relativistic fluid equations.
In \S\ref{sec:algorithm} we present discretization of the equations in
a general system of orthogonal curvilinear coordinates and describe
the numerical method. We include the details of the time
evolution algorithm based on characteristic tracing. The Riemann
solver is presented in \S\ref{sec:Riemann}. In \S\ref{sec:EoS} we 
consider different equation of states and introduce a 
new formulation suitable for relativistic regimes.
Several numerical tests are presented in \S\ref{sec:tests} and
conclusions are drawn in \S\ref{sec:concl}.

\section{THE EQUATIONS OF RELATIVISTIC HYDRODYNAMICS}\label{sec:equations}
%
%
%
%
%

In the framework of special relativity, the motion of an ideal fluid
is governed by the laws of particle number conservation and
energy-momentum conservation \citep{Lan_Lif59,Weinberg72}.
In the laboratory frame of reference, the conservation equations
written in divergence form are
\begin{equation} \label{eq:cl_1}
  \pd{}{t}\left(\begin{array}{c}
    D \\
   \A{m}  \\
   E \end{array}\right)  +
 \nabla\cdot \left( \begin{array}{c}
     D \A{v}		      \\
     \A{m} \A{v} + p\A{I}     \\
     \A{m}c^2 \end{array} \right) = 0  \,,
\end{equation}
where $\A{v}$, $D$, $\A{m}$, $E$ and $p$ define, respectively, the fluid 
three-velocity, density, momentum density, total energy density and pressure.
In the local rest frame the fluid can be described in terms of its
Lorentz-invariant thermodynamic quantities: the proper rest mass
density $\rho$, specific enthalpy $h$, and pressure $p$.
The transformation between the two frames of reference is given by
\begin{equation}\label{eq:UV1}
  \DS D	= \DS \gamma \rho  \,,
\end{equation}
\begin{equation}\label{eq:UV2}
 \DS \A{m} = \DS \rho h \gamma^2 \frac{\A{v}}{c^2} \,,
\end{equation}
\begin{equation}\label{eq:UV3}
  \DS E	= \DS \rho h \gamma^2 - p  \,,
\end{equation}
where $\gamma = \big(1 - \A{v}^2/c^2\big)^{-1/2}$ is the Lorentz
factor and $c$ is the speed of light. The specific enthalpy, $h$, is
related to the proper internal energy density, $e$, by
\begin{equation}
  h = \frac{e + p}{\rho} \,.
\end{equation}
An equation of state (EoS) provides an additional relation between
thermodynamic quantities and allows to close the system of
conservation laws (\ref{eq:cl_1}). Here we assume, without loss of
generality, that the EoS expresses the specific enthalpy $h$ as a 
function of the pressure $p$ and the specific proper volume 
$\tau=1/\rho$, i.e. $h=h(p,\tau)$.
This allows us to define the sound speed $c_s$ as
\begin{equation}\label{eq:sound_speed}
   \frac{c^2_s}{c^2} \equiv \left(\pd{p}{e}\right)_s  =
      \frac{\tau^2}{h}\pd{h}{\tau} \left(\pd{h}{p} - \tau\right)^{-1}
    \,,
\end{equation}
where the derivative in equation (\ref{eq:sound_speed}) has to be
taken at constant entropy $s$:
\begin{equation}
  dh|_s \equiv \left.\left(Tds + \frac{dp}{\rho}\right)\right|_s = \tau \, dp
  \,.
\end{equation}
We will consider only causal EoS, i.e., those for which $c_s < c$. 
For such equations of state, the hyperbolic property of equations 
(\ref{eq:cl_1}) is preserved \citep{Anile89}. 
In \S\ref{sec:EoS} we provide explicit expressions for
the sound speed for different EoS suitable for relativistic flows.

We describe the relativistic fluid in terms of the state vectors of
conservative, $\A{U} = (D,m_1,m_2,m_3,E)$, and primitive, $\A{V} =
(\rho,v_1,v_2,v_3,p)$, variables.  Here, $m_d$ and $v_d$ ($d=1,2,3$)
are the projection of the three-momentum $\A{m}$ and three-velocity
$\A{v}$ vectors along the coordinate axis, that is,
$m_d\equiv\nv_d\cdot\A{m}$ and $v_d \equiv\nv_d\cdot\A{v}$.

Equations (\ref{eq:UV1})--(\ref{eq:UV3}) define the map $\A{U} =
\A{U}(\A{V})$; the inverse relation gives $\A{V}$ in terms of
$\A{U}$:
\begin{equation}\label{eq:VU1}
   \rho                  =  \frac{D}{\gamma}  \,,
\end{equation}
\begin{equation}\label{eq:VU2}
   \frac{\A{v}}{c^2}  = \frac{\A{m}}{E + p}  \,,
\end{equation}
\begin{equation}\label{eq:VU3}
      p =  Dh\gamma - E   \,.
\end{equation}
This inverse map is not trivial due to the non-linearity introduced by
the Lorentz factor $\gamma$. Using equation (\ref{eq:VU2}) one can
express the Lorentz factor as a function of the pressure:
\begin{equation}\label{eq:gamma_p}
   \frac{1}{\gamma^2(p)} = 1 - \frac{\A{m}^2c^2}{\left(E + p\right)^2}
   \,.
\end{equation}
If $D$, $\A{m}$ and $E$ are given, equations (\ref{eq:VU3})
and (\ref{eq:gamma_p}) can be combined 
together to obtain an implicit expression for $p$:
\begin{equation}\label{eq:pressure_fun}
 f(p) = Dh\big(p,\tau(p)\big)\gamma(p) - E - p = 0
   \,.
\end{equation}
This equation must be solved numerically in order to recover
the pressure from a set of conservative variables $\A{U}$. 
Notice that $\tau = \tau(p)$ depends on the pressure $p$ 
through  $\tau = \gamma(p)/D$ and that 
the specific enthalpy $h$ is, in general, function of both $p$ and $\tau$,
$h = h(p,\tau(p))$.

If the Newton-Raphson method is used as the root finder for pressure,
the derivative of equation (\ref{eq:pressure_fun}) with respect to $p$
is needed. It can be expressed in terms of the derivatives of the
specific enthalpy , already appearing in equation
(\ref{eq:sound_speed}):
\begin{equation}
 \frac{df(p)}{dp} = D\gamma\pd{h}{p} - 
 \frac{\A{m}^2c^2\gamma^3}{(E + p)^3}\left(\gamma\pd{h}{\tau} + Dh\right) - 1
  \,.
\end{equation}
Equations (\ref{eq:VU1})--(\ref{eq:VU3}), together with the pressure
equation (\ref{eq:pressure_fun}), have to be solved at each time step
in order to provide the map $\A{V}(\A{U})$ needed in our numerical
method.

\section{DESCRIPTION OF THE ALGORITHM}\label{sec:algorithm}
%
%
%
%
%

Till now we considered the system of conservation laws
(\ref{eq:cl_1}]) written in divergence form. However, this
differential representation is valid as long as the flow remains
smooth. Since hyperbolic systems of conservation laws also admit
discontinuous solutions, we should consider the more general integral
form of the equations. We proceed as follows.

We introduce a generic system of orthogonal curvilinear coordinates,
$(x^1,x^2,x^3)$, and define $\nv_d$ to be the unit vectors associated
with the coordinate directions, $d=1,2,3$. Orthonormality requires
$\nv_d\cdot\nv_{d'} = 1$ when $d = d'$ and $\nv_d\cdot\nv_{d'} = 0$ 
otherwise. The geometrical properties are
then described in terms of the scale factors $h_1$, $h_2$ and $h_3$
which are, in general, functions of the coordinates, i.e., $h_d \equiv
h_d(x^1,x^2,x^3)$.  We divide our computational domain into control
volumes,
\begin{equation}\label{eq:volume}
   \Delta\vol_{ijk} = \int_{x^1_{i-\HALF}}^{x^1_{i+\HALF}}
                              \int_{x^2_{j-\HALF}}^{x^2_{j+\HALF}}
                              \int_{x^3_{k-\HALF}}^{x^3_{k+\HALF}}d\vol
  \,, \quad
   d\vol = h_1h_2h_3\,dx^1dx^2dx^3\,, 
\end{equation}
where the index triplet $(i,j,k)$ corresponds to the $\nv_1$, $\nv_2$ and
$\nv_3$ directions.  The integrals in equation (\ref{eq:volume}) are
evaluated between the lower ($x^1_{i-\HALF}, x^2_{j-\HALF},
x^3_{k-\HALF}$) and upper ($x^1_{i+\HALF}, x^2_{j+\HALF},
x^3_{k+\HALF}$) coordinate limits of the grid cell $ijk$, where
\begin{equation}
  x^1_{i+\HALF} - x^1_{i-\HALF} = \Delta x^1_i\,, \quad
  x^2_{j+\HALF} - x^2_{j-\HALF} = \Delta x^2_j\,, \quad
  x^3_{k+\HALF} - x^3_{k-\HALF} = \Delta x^3_k\,, \quad
\end{equation}
here $\Delta x^1_i$, $\Delta x^2_j$ and $\Delta x^3_k$ are the zone widths
in the three coordinate directions.

We restrict our attention to those coordinate systems for which the
volume defined by equation (\ref{eq:volume}) can be represented as the
product of three different functions, each one depending on one
coordinate only, i.e.,
\begin{equation}\label{eq:factorization}
  \Delta\vol_{ijk} = \left(\xi^1(x^1_{i+\HALF}) - \xi^1(x^1_{i-\HALF})\right)
                     \left(\xi^2(x^2_{j+\HALF}) - \xi^2(x^2_{j-\HALF})\right)
                     \left(\xi^3(x^3_{k+\HALF}) - \xi^3(x^3_{k-\HALF})\right)
\end{equation}

The three functions $\xi^1(x^1)$, $\xi^2(x^2)$, and $\xi^3(x^3)$ are the
generalized volume coordinates. 
These coordinates allow us to define the volumetric centroids $\xi^1_i$,
$\xi^2_j$, $\xi^3_k$ corresponding to the point $(x^1_i,
x^2_j, x^3_k)$ as
\begin{equation}\label{eq:vol_centroids_1}
  \xi^1_i \equiv \xi^1(x^1_i) = \frac{\xi^1(x^1_{i+\HALF}) + \xi^1(x^1_{i-\HALF})}{2} \,,
\end{equation}
\begin{equation}\label{eq:vol_centroids_2}
  \xi^2_j \equiv \xi^2(x^2_j) = \frac{\xi^2(x^2_{j+\HALF}) + \xi^2(x^2_{j-\HALF})}{2} \,,
\end{equation}
\begin{equation}\label{eq:vol_centroids_3}
  \xi^3_k \equiv \xi^3(x^3_k) = \frac{\xi^3(x^3_{k+\HALF}) + \xi^3(x^3_{k-\HALF})}{2} \,.
\end{equation}

Notice that in Cartesian coordinates, $(\xi^1, \xi^2, \xi^3) = (x,y,z)$,
and the volumetric centroids coincide with the geometrical zone
centers. In cylindrical geometry, $(x^1, x^2, x^3) = (r, \phi, z)$,
however,
\begin{equation}
  \xi^1(r)    = \frac{r^2}{2} \,, \quad
  \xi^2(\phi) = \phi          \,, \quad
  \xi^3(z)    = z             \,,
\end{equation}
with the locations corresponding to the volumetric centroids given by
\begin{equation}
 r_i    = \left(\frac{r^2_{i+\HALF} + r^2_{i-\HALF}}{2}\right)^{\HALF}\,, \quad
 \phi_j = \frac{\phi_{j+\HALF} + \phi_{j-\HALF}}{2}                   \,, \quad
 z_k    = \frac{z_{k+\HALF} + z_{k-\HALF}}{2}                         \,.
\end{equation}

Similarly, in spherical coordinates, $(x^1, x^2, x^3) = (r, \theta,
\phi)$, one obtains
\begin{equation}
  \xi^1(r)      = \frac{r^3}{3}  \,, \quad 
  \xi^2(\theta) = 1 - \cos\theta \,, \quad 
  \xi^3(\phi)   = \phi           \,,
\end{equation}
and the corresponding locations of the volumetric centroids are
\begin{equation}
  r_i      = \left(\frac{r^3_{i+\HALF} + r^3_{i-\HALF}}{2}\right)^{\frac{1}{3}}           \,, \quad
  \theta_j = \cos^{-1} \left(\frac{\cos\theta_{j+\HALF} + \cos\theta_{j-\HALF}}{2}\right) \,, \quad
  \phi_k   = \frac{\phi_{k+\HALF} + \phi_{k-\HALF}}{2}   \,.
\end{equation}

The areas of the zone faces are expressed by $A^d =\int \nv_d\cdot
d\A{A}$, where $d\A{A}$ is the area element $d\A{A}= dA^1\nv_1 +
dA^2\nv_2 + dA^3\nv_3$, and $dA^d\equiv d\vol/(h_d\,dx^d)$. The
result written in expanded form is
\begin{equation}
   A^1_{i\pm\HALF,j,k} = \int_{x^2_{j-\HALF}}^{x^2_{j+\HALF}}
                                      \int_{x^3_{k-\HALF}}^{x^3_{k+\HALF}}
       h_2(x^1_{i\pm\HALF},x^2,x^3)   h_3(x^1_{i\pm\HALF},x^2,x^3) \,dx^2 dx^3 \,,
\end{equation}
\begin{equation}
   A^2_{i,j\pm\HALF,k} = \int_{x^1_{i-\HALF}}^{x^1_{i+\HALF}}
                                      \int_{x^3_{k-\HALF}}^{x^3_{k+\HALF}}
       h_1(x^1,x^2_{j\pm\HALF},x^3)   h_3(x^1,x^2_{j\pm\HALF},x^3) \,dx^1 dx^3 \,,
\end{equation}
\begin{equation}
   A^3_{i,j,k\pm\HALF} = \int_{x^1_{i-\HALF}}^{x^1_{i+\HALF}}
                                      \int_{x^2_{j-\HALF}}^{x^2_{j+\HALF}}
      h_1(x^1,x^2,x^3_{k\pm\HALF})   h_2(x^1,x^2,x^3_{k\pm\HALF})\, dx^1 dx^2 \,.
\end{equation}
Explicit expressions for the scale factors and face areas are given in Appendix 
\ref{app:geo_1}.

Finally, the facial quadrature points located on the opposite bounding
faces,
\begin{equation} 
  \left(x^1_{i\pm\HALF}, x^2_j, x^3_k\right) \,, \quad 
  \left(x^1_i, x^2_{j\pm\HALF}, x^3_k\right) \,, \quad
  \left(x^1_i, x^2_j, x^3_{k\pm\HALF}\right) \,, \quad
\end{equation}
are used to compute the arc length separation,
\begin{equation}
  \Delta l^1_{ijk} = \int_{x^1_{i-\HALF}}^{x^1_{i+\HALF}}
                       h_1(x^1,x^2_j,x^3_k) dx^1   \,.
\end{equation}
Similar expressions can be given for $\Delta l^2_{ijk}$ and $\Delta
l^3_{ijk}$. 

The integral form of the equations can now be obtained by integrating
the system (\ref{eq:cl_1}) over a computational
volumes $\Delta \vol_{ijk}$ and over a discrete time interval $\Delta
t^n\equiv t^{n+1}-t^n$.  Using Gauss's theorem, the evolutionary
equations provide a relation between the volume averaged conserved
quantities and the time-averaged surface integrals for the divergence
terms:
\begin{equation}\label{eq:cl_2a}
  \Bar{D}^{n+1}_{ijk} - \bar{D}^n_{ijk} =
  - \frac{\Delta t^n}{\Delta\vol_{ijk}}\oint_{ijk}
      \left<D\A{v}\right>^n\cdot d\A{A} \,, 
\end{equation}
\begin{equation}\label{eq:cl_2b}
  \bar{m}^{n+1}_{d,ijk} - \bar{m}^n_{d,ijk} =
  - \frac{\Delta t^n}{\Delta\vol_{ijk}}
   \left[\oint_{ijk} \left<m_d\A{v}\right>^n\cdot d\A{A} +
   \int_{ijk} \nv_d\cdot\left<\nabla p + \A{S}_G\big(\A{m}\A{v}\big)\right>^n
    d\vol \right]   \,,
\end{equation}
\begin{equation}\label{eq:cl_2c}
  \bar{E}^{n+1}_{ijk} - \bar{E}^n_{ijk} = 
  - \frac{\Delta t^n}{\Delta\vol_{ijk}}\oint_{ijk}
    \left< \A{m}\right>^n\cdot d\A{A} \,,
\end{equation}
where $d = 1,2,3$ enumerate coordinate directions and the speed of light $c$ has
been set equal to unity, a convention to be 
used for the rest of this paper.
Equations (\ref{eq:cl_2a})--(\ref{eq:cl_2c}) express conservation of mass,
momentum, and energy.
In the equations above an overbar symbol denotes
a volume average,
\begin{equation}
  \bar{D}^n_{ijk} \equiv \frac{1}{\Delta\vol_{ijk}}\int_{ijk} D(x^1,x^2,x^3,t^n)\,d\vol \,,
\end{equation}
while a bracket denotes a time-averaged quantity,
\begin{equation}
  \left<D\A{v}\right>^n \equiv \frac{1}{\Delta t^n}\int_{t^n}^{t^{n+1}}
    D(x^1,x^2,x^3,t)\A{v}(x^1,x^2,x^3,t)\, dt \,.
\end{equation}
Notice that expressions (\ref{eq:cl_2a})--(\ref{eq:cl_2b}) are
exact and no approximation has been introduced so far.

The vector $\A{S}_G(\A{m}\A{v})$ accounts for
geometrical source terms arising from taking the divergence of the
dyad $\A{m}\A{v}$ in a curvilinear system of coordinates (explicit
expressions for the geometrical source terms are given in Appendix
\ref{app:geo_1}). 

\subsection{Evolution in Multi-dimensions}\label{sec:evolution}
%
%
%
%
%
%

Equations (\ref{eq:cl_2a})--(\ref{eq:cl_2c}) are integrated by first
solving the homogeneous problem, i.e. without the source term
$\A{S}_G$ in equation [\ref{eq:cl_2b}], separately from a source step.

Let $\A{\mathcal{H}}^{\Delta t}$ be the solution operator
corresponding to the homogeneous part of the problem and
$\A{\mathcal{S}}^{\Delta t}$ be the operator describing the contribution
of the source step over the time step $\Delta t$. Starting from the
initial data, $\bar{\A{U}}^n$, the solution for the next two time levels is
computed using operator splitting \citep{Strang68}:
\begin{equation}
 \bar{\A{U}}^{n+1} = \A{\mathcal{S}}^{\Delta t}\A{\mathcal{H}}^{\Delta t}
              \left(\bar{\A{U}}^n\right) \,,
\end{equation}
whereas
\begin{equation}
  \bar{\A{U}}^{n+2} = \A{\mathcal{H}}^{\Delta t}\A{\mathcal{S}}^{\Delta t}
              \left(\bar{\A{U}}^{n+1}\right) \,,
\end{equation}

Notice that this approach is second-order accurate in time, provided
that the two operators have at least the same order of accuracy and the
same time step is used for two consecutive levels.


The homogeneous operator $\A{\mathcal{H}}^{\Delta t}$ is based on the
spatially unsplit fully corner-coupled method
\citep[CTU,][]{Colella90,Saltzman94,MC01,MC02}.  The conservative
update results from acting $\A{\mathcal{H}}^{\Delta t}$ on
$\bar{\A{U}}^{n}$:
\begin{equation}\label{eq:unsplit}
  \A{\mathcal{H}}^{\Delta t}\left(\bar{\A{U}}^{n}\right)  = 
    \bar{\A{U}}^{n}  
  + \A{\mathcal{L}}^{1}\left(\A{U}^{n+\HALF}_{i+\HALF}, \A{U}^{n+\HALF}_{i-\HALF}\right) 
  + \A{\mathcal{L}}^{2}\left(\A{U}^{n+\HALF}_{j+\HALF}, \A{U}^{n+\HALF}_{j-\HALF}\right) 
  + \A{\mathcal{L}}^{3}\left(\A{U}^{n+\HALF}_{k+\HALF}, \A{U}^{n+\HALF}_{k-\HALF}\right) \,,
\end{equation}
where $\bar{\A{U}}^n = \left( \bar{D}_{ijk}, \bar{\A{m}}_{ijk},
\bar{E}_{ijk} \right)^n$ is a state vector of volume-averaged
conserved quantities at time $t=t^n$.  Here we adopted the convention in which the
integer-valued subscripts $i,j,k$ are omitted when referring to
three-dimensional quantities while the half integer values are used to
denote zone edges. The same convention is used throughout the rest of the
paper.

The last three terms in equation (\ref{eq:unsplit}), $\A{\mathcal{L}}^d$,
represent one-dimensional flux-differencing operators, where $d=1,2,3$
labels directions. For $d=1$ (expressions for $d=2$ and $d=3$ are
obtained by suitable permutations of indices) one has
\begin{equation}\label{eq:1d_operator}\begin{split}
  \A{\mathcal{L}}^{1}\left(\A{U}^{n+\HALF}_{i+\HALF},\A{U}^{n+\HALF}_{i-\HALF}\right)
                             = &
  - \frac{\Delta t}{\Delta\vol}\left[
 \A{F}^{1}\left(\A{U}^{n+\HALF}_{i+\HALF}\right)A^{1}_{i+\HALF}
-\A{F}^{1}\left(\A{U}^{n+\HALF}_{i-\HALF}\right)A^{1}_{i-\HALF}
   \right] \\ 
 & - \frac{\Delta t}{\Delta l^1} \left[
     \A{P}^1\left(\A{U}^{n+\HALF}_{i+\HALF}\right)
   - \A{P}^1\left(\A{U}^{n+\HALF}_{i-\HALF}\right)\right]   \,,
\end{split}\end{equation}
where
\begin{equation}
 \A{F}^1(\A{U}) = \left(\begin{array}{c}
       Dv_1    \\
       m_1v_1  \\
       m_2v_1  \\
       m_3v_1  \\
       m_1   \end{array} \right) \,,
\end{equation}
is the flux vector, and
\begin{equation}
   \A{P}^1(\A{U}) = \left(\begin{array}{c}
      0   \\
      p   \\
      0   \\
      0   \\
      0     \end{array}\right)  \,,
\end{equation}
is the pressure term. The state vectors $\A{U}^{n+\HALF}_{i\pm\HALF}$
are solutions to Riemann problems with suitable time-centered 
left and right states properly computed at each facial quadrature point.
For example,
\begin{equation}\label{eq:riemann3f}
 \A{U}^{n+\HALF}_{i+\HALF} = 
 \Riemann \left( \A{U}^{n+\HALF}_{i+\HALF,L}, 
              \A{U}^{n+\HALF}_{i+\HALF,R}\right) \,,
\end{equation}
where $\Riemann(\cdot,\cdot)$ denotes the solution to the Riemann
problem (see \S\ref{sec:Riemann} for details).

In the three-dimensional case, the conservative update
(\ref{eq:unsplit}) involves the following four steps. In the first
predictor step we construct initial left and right states just as in
the one-dimensional case. In the second predictor step the initial
states are corrected for contributions from transverse directions to
form secondary predictors. These secondary predictors are used in the
third step to calculate the fully corner coupled states. Finally, the
fluxes required by the conservative update follow from the solutions
of Riemann problems with the fully corner-coupled states used as
input. In what follows we present the details of this procedure.

The starting point for the construction of the input left and right
states to the Riemann problems is the predictor step described in
\S\ref{sec:1d_solver} below. In this step we use Taylor
expansion to obtain edge- and time-centered estimates for the left and
right states $\hat{\A{U}}^{n+\HALF}_{i+\HALF,L}$ and
$\hat{\A{U}}^{n+\HALF}_{i+\HALF,R}$.

Next the following 6 secondary predictors are calculated:
\begin{equation}\label{eq:predictor_1}
 \left.\begin{array}{c}
\DS  \A{U}^{'1}_{j\pm\HALF,S} = \hat{\A{U}}^{n+\HALF}_{j\pm\HALF,S}   \\  \noalign{\medskip}
\DS  \A{U}^{'1}_{k\pm\HALF,S} = \hat{\A{U}}^{n+\HALF}_{k\pm\HALF,S} 
   \end{array}\right\}
   + \frac{1}{3}\A{\mathcal{L}}^1\left(\hat{\A{U}}^{n+\HALF}_{i+\HALF},
                                       \hat{\A{U}}^{n+\HALF}_{i-\HALF}
  \right) \,,
\end{equation}
\begin{equation}\label{eq:predictor_2}
 \left.\begin{array}{c}
\DS  \A{U}^{'2}_{i\pm\HALF,S} = \hat{\A{U}}^{n+\HALF}_{i\pm\HALF,S}   \\ \noalign{\medskip}
\DS  \A{U}^{'2}_{k\pm\HALF,S} = \hat{\A{U}}^{n+\HALF}_{k\pm\HALF,S} 
   \end{array}\right\}
   + \frac{1}{3}\A{\mathcal{L}}^2\left(\hat{\A{U}}^{n+\HALF}_{j+\HALF},
                                       \hat{\A{U}}^{n+\HALF}_{j-\HALF}
  \right) \,,
\end{equation}
\begin{equation}\label{eq:predictor_3}
 \left.\begin{array}{c}
\DS  \A{U}^{'3}_{i\pm\HALF,S} =  \hat{\A{U}}^{n+\HALF}_{i\pm\HALF,S}  \\ \noalign{\medskip}
\DS  \A{U}^{'3}_{j\pm\HALF,S} =  \hat{\A{U}}^{n+\HALF}_{j\pm\HALF,S} 
   \end{array}\right\}
   + \frac{1}{3}\A{\mathcal{L}}^3\left(\hat{\A{U}}^{n+\HALF}_{k+\HALF},
                                       \hat{\A{U}}^{n+\HALF}_{k-\HALF}
  \right) \,,
\end{equation}
where we adopt the convention that $S=L$ at $i+\HALF$ and $S=R$ at
$i-\HALF$ (similar formulae are obtained for $j\pm\HALF$,
$k\pm\HALF$).

The flux differencing operators on the right hand side of
equations (\ref{eq:predictor_1})--(\ref{eq:predictor_3}) 
are obtained by solving Riemann problems with the
states obtained in the first corrector step, e.g.,
\begin{equation}\label{eq:riemann3i}
 \hat{\A{U}}^{n+\HALF}_{i+\HALF}  = 
   \Riemann\left(\hat{\A{U}}^{n+\HALF}_{i+\HALF,L},
                    \hat{\A{U}}^{n+\HALF}_{i+\HALF,R}\right)\,.
\end{equation}

In the third step we obtain the fully corner-coupled states,
\begin{equation}\label{eq:corrector_1}
  \A{U}^{n+\HALF}_{i\pm\HALF,S} = 
     \hat{\A{U}}^{n+\HALF}_{i\pm\HALF,S}  + \frac{1}{2}\left[
       \A{\mathcal{L}}^{2}\left(\A{U}^{'3}_{j+\HALF}, \A{U}^{'3}_{j-\HALF}\right) 
 +     \A{\mathcal{L}}^{3}\left(\A{U}^{'2}_{k+\HALF}, \A{U}^{'2}_{k-\HALF}\right)\right] \,,
\end{equation}
\begin{equation}\label{eq:corrector_2}
  \A{U}^{n+\HALF}_{j\pm\HALF,S} = 
    \hat{\A{U}}^{n+\HALF}_{j\pm\HALF,S} + \frac{1}{2}\left[
      \A{\mathcal{L}}^{3}\left(\A{U}^{'1}_{k+\HALF}, \A{U}^{'1}_{k-\HALF}\right)
    + \A{\mathcal{L}}^{1}\left(\A{U}^{'3}_{i+\HALF}, \A{U}^{'3}_{i-\HALF}\right)\right] \,,
\end{equation}
\begin{equation}\label{eq:corrector_3}
  \A{U}^{n+\HALF}_{k\pm\HALF,S} = 
    \hat{\A{U}}^{n+\HALF}_{k\pm\HALF,S} + \frac{1}{2}\left[
       \A{\mathcal{L}}^{1}\left(\A{U}^{'2}_{i+\HALF}, \A{U}^{'2}_{i-\HALF}\right)
    +  \A{\mathcal{L}}^{2}\left(\A{U}^{'1}_{j+\HALF}, \A{U}^{'1}_{j-\HALF}\right)\right] \,.
\end{equation}

As in the previous step, quantities appearing in the flux
differencing follow the solution of appropriate Riemann problems. For
example,
\begin{equation}\label{eq:riemann6i}
 \A{U}^{'1}_{j+\HALF} = 
 \Riemann\left( \A{U}^{'1}_{j+\HALF, L},
                \A{U}^{'1}_{j+\HALF, R}\right) \,.
\end{equation}

The final step is the conservative update in which we use the fluxes obtained
by solving Riemann problems with the fully corner-coupled states
(eq. [\ref{eq:corrector_1}]--[\ref{eq:corrector_3}]).

In three dimensions a total of 12 Riemann problems are required:
3 to obtain the secondary predictors 
(eq. [\ref{eq:predictor_1}]--[\ref{eq:predictor_3}]), 6 to obtain
the fully corner-coupled states 
(eq. [\ref{eq:corrector_1}]--[\ref{eq:corrector_3}]), and 3 necessary 
for the conservative update (eq. [\ref{eq:unsplit}]).
The six secondary predictors, equations (\ref{eq:predictor_1})--(\ref{eq:predictor_3}), 
are not required in two dimensions and the fully corner-coupled states
become 
\begin{equation}\label{eq:corrector2d_1}
  \A{U}^{n+\HALF}_{i\pm\HALF,S} = 
     \hat{\A{U}}^{n+\HALF}_{i\pm\HALF,S}  + \frac{1}{2}
       \A{\mathcal{L}}^{2}\left(\hat{\A{U}}^{n+\HALF}_{j+\HALF}, \hat{\A{U}}^{n+\HALF}_{j-\HALF}\right)\,,
\end{equation}
\begin{equation}\label{eq:corrector2d_2}
  \A{U}^{n+\HALF}_{j\pm\HALF,S} = 
     \hat{\A{U}}^{n+\HALF}_{j\pm\HALF,S}  + \frac{1}{2}
       \A{\mathcal{L}}^{1}\left(\hat{\A{U}}^{n+\HALF}_{i+\HALF}, \hat{\A{U}}^{n+\HALF}_{i-\HALF}\right)\,.
\end{equation}

Therefore, in 2D, the conservative update involves the solution of 4
Riemann problems: 2 in the predictor step and 2 in the corrector
step. 

Contribution from geometrical source terms is included by solving the
ordinary differential equation
\begin{equation}\label{eq:source_ode}
 \frac{d\A{U}}{dt} = \A{S}_G(\A{U}) \,,
\end{equation}

with the solution from the previous step (either advection or possibly
source term calculation if Strang splitting is used) used as the
initial data. The solution is obtained with a second-order Runge-Kutta
algorithm, and can be written in operator form as
\begin{equation}
 \begin{array}{c}
  \DS \A{\mathcal{S}}^{\Delta t}\left(\A{U}\right) = \frac{1}{2}\left[\A{U} +
                  \A{U}^* + \Delta t\A{S}_G\left(\A{U}^*\right)\right]  \,,  \\ \noalign{\medskip}
  \DS \A{U}^* = \A{U} + \Delta t\A{S}_G\left(\A{U}\right) \, .
 \end{array}
\end{equation}
 
Finally, the choice of the time step $\Delta t$ is based on the
Courant-Friederichs-Lewy (CFL) condition \citep{CFL28}:

\begin{equation}
  \Delta t = C_{a}\times \min_{i,j,k}\left(
   \frac{ \Delta l_{ijk}^1}{|\lambda_{\max}^1|},
   \frac{ \Delta l_{ijk}^2}{|\lambda_{\max}^2|},
   \frac{ \Delta l_{ijk}^3}{|\lambda_{\max}^3|}\right) \,.
\end{equation}
Here $\lambda_{\max}^d$ is the fastest wave speed in the $\nv_d$
direction, and $0 < C_{a} < 1$ is the limiting factor.

\subsection{The predictor step}\label{sec:1d_solver}
%
%
%
%
%

In what follows we provide a detailed description of the predictor
step which constitutes the first step in the multi-stage procedure
described above. For simplicity we describe the operator applied 
to the first coordinate direction; modifications required for other
directions are straightforward. The predictor step is comprised of the
reconstruction algorithm followed by the calculation of left and right
states at the zone interfaces. These states are used as input data to
the Riemann problems in the second predictor step (eq. 
[\ref{eq:predictor_1}]--[\ref{eq:predictor_3}]) and
are required in the construction of the fully corner coupled 
states, equations (\ref{eq:corrector_1})--(\ref{eq:corrector_3}) in
3-D and equations (\ref{eq:corrector2d_1}), (\ref{eq:corrector2d_2}) 
in 2-D. Throughout this section we make use of the following
abbreviations: $x=x^1$, $\xi=\xi^1$, $h=h^1$ and $A = A^1$.

During the reconstruction step a quartic polynomial is passed
through the volume-averaged values $\bar{\A{U}}^n_i$. 
The polynomial provides interface values for the hydrodynamic 
variables, $\A{V}^n_{i+\HALF,S}$ ($S=L,R$) which are fourth-order 
accurate in space in smooth regions of the flow. 
The interpolation may be done either in the volume coordinate $\xi$ 
(CW84) or in the spatial coordinate $x$ \citep{BL93}.
Here we follow the former prescription and the details are given in
Appendix \ref{app:PPM_curv}.

Once the interpolated zone-edge values are provided, 
we define a one-dimensional piece-wise parabolic profile 
inside each cell $i$: 
\begin{equation}\label{eq:parabola}
  \A{V}^n(\xi) = \A{V}^n_{i-\HALF,R} + \frac{\xi - \xi_{i-\HALF}}{\Delta \xi_i}
  \left[
      \Delta \A{V}^n_i + \A{V}^n_{6i}\left(1 - 
      \frac{\xi - \xi_{i-\HALF}}{\Delta \xi_i} \right)
  \right]  \quad  \textrm{for} \quad \xi_{i-\HALF} \le \xi \le \xi_{i+\HALF}  \,,
\end{equation}
where 
\begin{equation}
  \Delta\A{V}^n_i = \A{V}^n_{i+\HALF,L} - \A{V}^n_{i-\HALF,R}\,, \quad
  \A{V}_{6i}^n = 6\left(\A{V}^n_i - 
          \frac{\A{V}^n_{i+\HALF,L} + \A{V}^n_{i-\HALF,R}}{2}\right) \,.
\end{equation}

Notice, however, that when the interpolation is done in the spatial coordinate
$x$ rather than in the volume coordinate $\xi$,
equation (\ref{eq:parabola}) has to be modified by
replacing $\xi(x)$ with $x$. Moreover, the definition of $\A{V}_{6i}$
becomes geometry dependent \citep{BL93}.

Second-order accuracy in time is obtained by considering the
one-dimensional relativistic equations in quasi-linear
form \citep[derived for the one-dimensional case by][]{MM96}:
\begin{equation} \nonumber
  \pd{\A{V}}{t} + \frac{1}{h}\A{A}(\A{V})\cdot\pd{\A{V}}{x} =
      \A{S}(\A{V})\,.
\end{equation}
Here
\begin{equation}
   \A{V} = \left(\begin{array}{c}
      \rho \\
       v_1  \\
       v_2  \\
       v_3  \\
       p   \end{array}\right) \,, \quad
  \A{S}(\A{V}) =
  \frac{v_1}{\Delta^2}\pd{}{\xi}\left(\frac{1}{h}\pd{\xi}{x}\right)\left(\begin{array}{c}
     -\rho   \\ 
     v_1c_s^2/\gamma^2 \\
     v_2c_s^2/\gamma^2 \\
     v_3c_s^2/\gamma^2 \\
     -\rho hc_s^2 \end{array}\right) \,,
\end{equation}
are, respectively, the vector of primitive variables and geometrical 
sources, while
\begin{equation}\label{eq:1d-prim}
   \A{A}(\A{V}) =   \frac{1}{\Delta^2} \left( \begin{array}{ccccc}
 v_1\Delta^2 &	\rho	 &  0	 &   0	 &
   \DS  -\frac{v_1}{h\gamma^2} \\\noalign{\medskip}
 0 &  v_1(1-c_s^2) & 0 & 0 &
   \DS    \frac{\eta^2}{\rho h\gamma^2} \\\noalign{\medskip}
 0 & \DS -\frac{v_2 c_s^2}{\gamma^2 } &  v_1\Delta^2 & 0 &
   \DS  - \frac{v_1 v_2 \left(1 - c_s^2\right)}{\rho h\gamma^2 } \\\noalign{\medskip}
 0 & \DS -\frac{v_3 c_s^2}{\gamma^2} & 0 & v_1\Delta^2 &
   \DS  - \frac{v_1 v_3 \left(1 - c_s^2\right)}{\rho h\gamma^2 } \\\noalign{\medskip}
 0 & \rho h c_s^2 & 0 & 0 & v_1(1 - c_s^2)
      \end{array}\right)\,,
\end{equation}
is the Jacobian, $\Delta^2 = 1 - \A{v}^2c_s^2$ and $\eta^2 = 1 -
v_1^2 - c_s^2 (v_2^2+v_3^2)$.

The left and right states $\hat{\A{V}}^{n+\HALF}_{i+\HALF,L}$ and 
$\hat{\A{V}}^{n+\HALF}_{i-\HALF,R}$, are computed using the
characteristic information carried to the zone edges $\xi_{i\pm\HALF}$
by each family of waves emanating from the zone center during a single
time step $\Delta t$. Firstly, for each characteristic $\#$ we find the
domain of dependence for each interface:
\begin{equation}\label{eq:domains}
  \xi^{\#}_{i-\HALF,R} = \xi\left(x_{i-\HALF} + \Delta x^\#_{i-\HALF} \right) \,, 
  \quad
  \xi^{\#}_{i+\HALF,L} = \xi\left(x_{i+\HALF} - \Delta x^\#_{i+\HALF} \right) \,,
\end{equation}
where $\Delta x^\#_{i-\HALF}$ and $\Delta x^\#_{i+\HALF}$ are defined through 
\begin{equation}
  \int_{x_{i-\HALF}}^{x_{i-\HALF} + \Delta x^\#_{i-\HALF}} h(x,x^2_j, x^3_k)dx =
           \Delta t\,\max(0,-\lambda^{\#}_i) \,,
\end{equation}
\begin{equation}
  \int_{x_{i+\HALF} - \Delta x^\#_{i+\HALF}}^{x_{i+\HALF}} h(x,x^2_j, x^3_k)dx =
           \Delta t\,\max(0, \lambda^{\#}_i)\,.
\end{equation}
Here $\# = \{+, 0^{(1,2,3)}, -\}$ labels the
characteristic field, and $\lambda^{\#}$ is the corresponding 
eigenvalue \citep{FK96,DFIM98},
\begin{equation}\label{eq:eigenvalues}
  \lambda^{-} = \frac{\gamma v_1\left(1-c_s^2\right) -
	     c_s\eta}{\gamma\Delta^2}  \,, \quad
  \lambda^{0^{(1,2,3)}} = v_1      \,, \quad
  \lambda^{+} = \frac{\gamma v_1\left(1-c_s^2\right) +
	     c_s\eta}{\gamma\Delta^2}\,.
\end{equation}

As in classical hydrodynamics, the characteristics
consist of two genuinely non-linear fields $\# = +, -$ and three
linearly degenerate fields $\# = 0^{(1,2,3)}$. Also, note that in
cylindrical or spherical coordinates, $h \equiv h^1(x^2_j,x^3_k)$ does
not depend on $x$ and therefore $\Delta x^\#_{i\pm\HALF} = \Delta
t\,\max(0,\pm\lambda^\#_i)/h(x^2_j,x^3_k)$.

Next, we calculate the average of $\A{V}(\xi)$ over the
domain of dependence lying to the left (for
$\hat{\A{V}}^{n+\HALF}_{i+\HALF,L}$) or to the right (for
$\hat{\A{V}}^{n+\HALF}_{i-\HALF,R}$) of the interface:
\begin{equation}\label{eq:averages}
  \A{V}^{\#}_{i+\HALF,L} = \frac{1}{\xi_{i+\HALF} - \xi^{\#}_{i+\HALF,L}}
                            \int^{\xi_{i+\HALF}}_{\xi^{\#}_{i+\HALF,L}}
                            \A{V}^n(\xi) d\xi \,,  \quad
  \A{V}^{\#}_{i-\HALF,R} = \frac{1}{\xi^{\#}_{i-\HALF,R} - \xi_{i-\HALF}}
                          \int^{\xi^{\#}_{i-\HALF,R}}_{\xi_{i-\HALF}}
                           \A{V}^n(\xi) d\xi\,.
\end{equation}

The zone averages are obtained by straightforward integration of
the parabolic zone profiles (eq. [\ref{eq:parabola}]):
\begin{equation}\label{eq:states_L}
  \A{V}^{\#}_{i+\HALF,L} = \A{V}^n_{i+\HALF,L} -
     \frac{\xi_{i+\HALF} - \xi^{\#}_{i+\HALF,L}}{2\Delta \xi_i}\left[
      \Delta \A{V}^n_i- \A{V}^n_{6i}\left(1 - \frac{2}{3}
      \frac{\xi_{i+\HALF} - \xi^{\#}_{i+\HALF,L}}{\Delta \xi_i} \right)
      \right]  \,,
\end{equation}
\begin{equation}\label{eq:states_R}
  \A{V}^{\#}_{i-\HALF,R} = \A{V}^n_{i-\HALF,R} +
      \frac{\xi^{\#}_{i-\HALF,R} - \xi_{i-\HALF}}{2\Delta \xi_i}\left[
      \Delta \A{V}^n_i + \A{V}^n_{6i}\left(1 - \frac{2}{3}
      \frac{\xi^{\#}_{i-\HALF,R} - \xi_{i-\HALF}}{\Delta \xi_i} \right)
        \right]\,.
\end{equation}

Once the integrals are obtained, we use upwind limiting to select
only those characteristics which contribute to the effective left and
right states \citep{CW84,MC02}. This requires expanding the matrix
$\A{A}(\A{V})$ (eq. [\ref{eq:1d-prim}]) in terms of its eigenvalues and
left and right eigenvectors.  Specifically, upwind limiting for the
left states consists in selecting those characteristics with positive
speeds (i.e. emanating from the cell center $\xi_i$ towards the zone
interface). Similarly, for the right states we consider only
characteristics with negative speeds. The final result is:
\begin{equation}\label{eq:riemann_state_L}
 \hat{\A{V}}^{n+\HALF}_{i+\HALF,L} = \A{V}^+_{i+\HALF,L} -
      \sum_{\lambda^{\#}_i>0}^{} \left[\A{l}_i^{\#} \cdot  
	\left(\A{V}^+_{i+\HALF,L} - \A{V}^{\#}_{i+\HALF,L} 
	\right)\right] \A{r}_i^{\#}  + 
    \frac{\Delta t}{2} \A{S}_i  \,,
\end{equation}
\begin{equation}\label{eq:riemann_state_R}
 \hat{\A{V}}^{n+\HALF}_{i-\HALF,R} = \A{V}^-_{i-\HALF,R} -
      \sum_{\lambda^{\#}_i<0} \left[\A{l}_i^{\#}  \cdot
      \left(\A{V}^-_{i-\HALF,R} - \A{V}^{\#}_{i-\HALF,R} 
	    \right) \right]\A{r}_i^{\#}  +
    \frac{\Delta t}{2} \A{S}_i \,.
\end{equation}
Here $\A{l}^{\#}$ and $\A{r}^{\#}$ are the left and right bi-orthonormal
eigenvectors of $\A{A}(\A{V})$:
%
%
\begin{equation}\label{eq:left_eigenvct}
 \A{l}^\pm = \left(\begin{array}{c}
     0	\\  \noalign{\medskip}
 \DS \pm\,{\frac {\rho\,\gamma}{2c_s\eta}}   \\  \noalign{\medskip}
     0 \\  \noalign{\medskip}
     0 \\  \noalign{\medskip}
  \DS \frac{1}{2hc_s^2} \\ \noalign{\medskip}  
 \end{array}\right)	 \,, \quad
 \A{l}^{0^{(1,2,3)}} = \left(\begin{array}{c}
   1  \\  \noalign{\medskip}
   0  \\  \noalign{\medskip}
   0  \\  \noalign{\medskip}
   0  \\  \noalign{\medskip}
  \DS  -\frac{1}{hc_s^2} \\ \noalign{\medskip} 
\end{array}\right)	\, , \quad
 \left(\begin{array}{c}
   0  \\  \noalign{\medskip}
 \DS     \frac{v_1v_2}{1-v_1^2} \\  \noalign{\medskip}
   1  \\  \noalign{\medskip}
   0  \\  \noalign{\medskip}
  \DS \frac {v_2}{\gamma^2 \rho h\left(1 - v_1^2\right)} \\ \noalign{\medskip} 
\end{array}\right)	\, , \quad
  \left(\begin{array}{c}
   0  \\  \noalign{\medskip}
 \DS    \frac{v_1v_3}{1-v_1^2} \\  \noalign{\medskip}
   0  \\  \noalign{\medskip}
   1  \\  \noalign{\medskip}
  \DS  \frac {v_3}{\gamma^2 \rho h\left(1 - v_1^2\right)} \\  \noalign{\medskip}
\end{array}\right) \,,
\end{equation}
%
%
\begin{equation}\label{eq:right_eigenvct}
  \A{r}^\pm = \left(\begin{array}{c}
	   1 \\        \noalign{\medskip}
 \DS \pm{\frac {c_s\eta}{\rho\,\gamma}} \\ \noalign{\medskip}
 \DS \mp\frac{c_s v_2\left(\gamma\eta v_1 \pm c_s\right)}
                                   {\gamma^2\rho\left(1 - v_1^2\right)} \\  \noalign{\medskip}
 \DS  \mp\frac{c_s v_3\left(\gamma\eta v_1 \pm c_s\right)}
                                   {\gamma^2\rho\left(1 - v_1^2\right)} \\  \noalign{\medskip}
   hc_s^2  \end{array} \right)\,, \quad
 \A{r}^{0^{(1,2,3)}} = \left(\begin{array}{c}
  1 \\	\noalign{\medskip}
  0 \\	\noalign{\medskip}
  0 \\	\noalign{\medskip}
  0 \\	\noalign{\medskip}
  0 \\	
 \end{array} \right) \,, \quad
 \left(\begin{array}{c}
  0 \\	\noalign{\medskip}
  0 \\	\noalign{\medskip}
  1 \\	\noalign{\medskip}
  0 \\	\noalign{\medskip}
  0 \\	\noalign{\medskip}
 \end{array} \right) \,, \quad
  \left(\begin{array}{c}
  0 \\	\noalign{\medskip}
  0 \\	\noalign{\medskip}
  0 \\	\noalign{\medskip}
  1 \\	\noalign{\medskip}
  0 \\	\noalign{\medskip}
 \end{array} \right) \,.
\end{equation}

It can be verified that left and right eigenvectors are properly
normalized, so that $\A{r}^{\#}\cdot \A{l}^{\#'} = 1$ for $\#=\#'$ and
$\A{r}^{\#}\cdot \A{l}^{\#'} = 0$ otherwise.
Notice that, in the limit of vanishing transverse velocities, 
expressions (\ref{eq:left_eigenvct}) and (\ref{eq:right_eigenvct})
reduce to the one-dimensional left and right eigenvectors given 
by \cite{MM96} (apart from the normalization factor $\rho\gamma^2/c_s$
int $\A{r}^\pm$ and $\A{l}^\pm$).

\subsection{Solution of the Riemann Problem}\label{sec:Riemann}
%
%
%
%
%

The Riemann solver describes the evolution of a discontinuity
separating two arbitrary constant hydrodynamic states, $\A{V}_L$ and
$\A{V}_R$:
\begin{equation}\label{eq:riemann_problem}
 \A{V}(\A{x},t=0) = \left\{\begin{array}{ll}
	      \A{V}_L	& \qquad \textrm{for}\quad x<0 \,,\\
	      \A{V}_R	& \qquad \textrm{for}\quad x>0 \,.
	 \end{array}\right.
\end{equation}

\begin{figure}
 \epsscale{0.65}
 \plotone{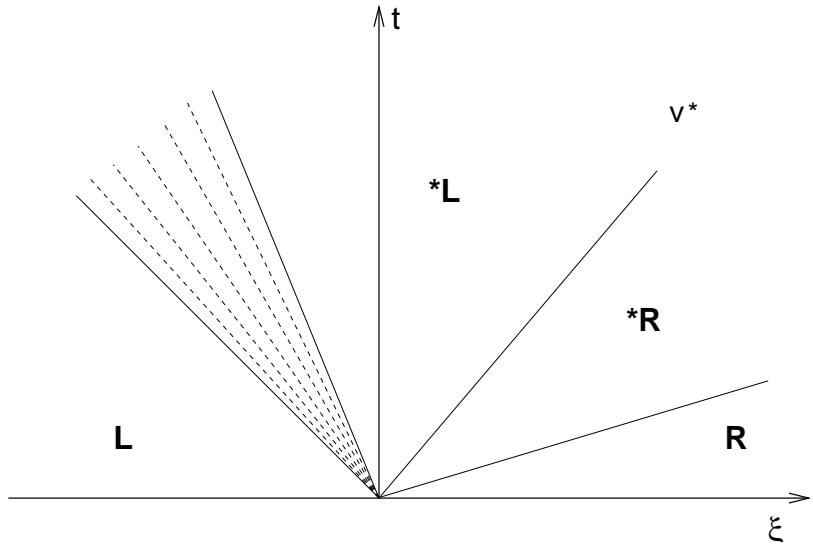}
 \caption{\footnotesize Solution of the Riemann problem in the $\xi-t$ plane. 
          The discontinuity between the two states $L$ and $R$ initially
          located at $\xi=0$ decays into a set of three waves. The emerging 
          pattern divides the $\xi-t$ plane into the four distinct regions
          labeled by $L, L^*, R^*, R$. A contact discontinuity separates state
          $L^*$ from state $R^*$, while the $R^*R$ and $LL^*$ waves can be either 
          shocks or rarefactions (in this picture the $LL^*$ and $R^*R$ wave
          are, respectively, a rarefaction wave and a shock).\label{fig:riemann_fan}}
\end{figure}

Like in the Newtonian case, the solution is self-similar (i.e. it is a
function of $x/t$ alone), and the decay of the initial discontinuity
gives rise to a three-wave pattern (see Fig. \ref{fig:riemann_fan}):
a contact discontinuity separating
two non-linear waves, either shocks or rarefactions. The contact
discontinuity moves at the fluid speed and both the normal velocity
$v_1$ and pressure $p$ are continuous across it, while density $\rho$
and tangential velocities $v_{2,3}$ generally experience jumps.
Across a shock wave, however, all the components of $\A{V}$ can be
discontinuous and their values ahead and behind the shock are related
by the Rankine-Hugoniot jump conditions.  Inside a rarefaction wave,
on the contrary, density, velocity and pressure have smooth profiles
given by a self-similar solution of the flow equations. In the
multidimensional case, the relation between the hydrodynamical states
at the head and the tail of the rarefaction wave can be cast in one
non-linear ordinary differential equation \citep{PMM00}
the solution of which is usually computationally expensive. Therefore
our solver relies on the ``two-shock'' approximation: a Riemann
problem is solved at each zone interface by approximating rarefaction
waves as shocks. This approach has been extensively used in
Godunov-type codes \citep[e.g.,][]{CW84,WC84,DW97}.

The jump conditions for the left ($S=L$) and right-going shock ($S=R$)
can be written as \citep{Taub48, PMM00}:
{\setlength\arraycolsep{1.5pt} \begin{eqnarray}
\label{eq:rankine1}  \left[\frac{1}{D}\right]  = -\zeta\big[v_1\big] \,, \\ \nonumber \\
\label{eq:rankine2}  \left[h \gamma v_1\right] = \zeta\big[p\big]    \,, \\ \nonumber \\
\label{eq:rankine3}  \left[h \gamma v_2\right] = 0                   \,, \\ \nonumber \\
\label{eq:rankine4}  \left[h \gamma v_3\right] = 0                   \,, \\ \nonumber \\
\label{eq:rankine5}  \left[h\gamma - \frac{p}{D}\right] = \zeta\big[pv_1\big]\,,
\end{eqnarray}}
where $\left[q\right]=q-q_S$ denotes the difference between the
pre-shock ($q_S$, $S=L,R$) and the post-shock state $q$,
$\zeta = \gamma_{sh}/j$, $\gamma_{sh}=(1-v_{sh}^2)^{-1/2}$ is the Lorentz 
factor of the shock, $v_{sh}$ is the shock velocity and $j$ is the mass
flux across the shock. 

We note that in relativistic flows the tangential velocities can 
exhibit jumps across the shock \citep{PMM00, RZP03}.
This is one of the major differences from the classical case caused on
one hand by the coupling between different velocity components
through the Lorentz factor (kinematically relativistic regime where
the flow speeds become comparable to the speed of light) and on the
other hand by the specific enthalpy being tied to the Lorentz factor
(thermodynamically relativistic regime when the pressure $p$ is
comparable to or greater than the rest mass energy density $\rho$).
Both characteristics are absent from Newtonian hydrodynamics.

The solution of the Riemann problem is obtained as follows. First, we
combine equations (\ref{eq:rankine1})--(\ref{eq:rankine5}) to obtain
the Taub adiabat \citep{Taub48}, the relativistic version of the Hugoniot shock
adiabat. The adiabat relates thermodynamic quantities -- enthalpy
$h$, pressure $p$, and specific proper volume $\tau$ -- ahead and
behind the shock:
\begin{equation}\label{eq:taub}
 \left[h^2\right] = \big(h\tau + h_S\tau_S\big)\left[p\right] \,.
\end{equation}

The above equation together with an EoS, $h=h(p,\tau)$, is solved to
obtain $\tau$ and $h$ as functions of the post-shock pressure
$p$ (explicit solutions for selected equations of state are
provided in \S\ref{sec:EoS}). These solutions together with
the definition of the mass flux,
\begin{equation}\label{eq:mass_flux_def}
  j^2 \equiv \gamma^2_{sh}D^2_S(v_{sh} - v_S)^2\,,
\end{equation}
lead to a quadratic equation for $\zeta$. That equation can be solved
to obtain
\begin{equation}\label{eq:zeta}
    \zeta_{\pm}(p,S) = \frac{V_S v_S \pm \sqrt{V_S^2 + (1 - v_S^2)/j^2(p,S)}}
		       {1 - v_S^2} \,,
\end{equation}
where $V_S = 1/D_S$, and for the mass flux we use the 
expression \citep{MM94,MM96,PMM00}
\begin{equation}\label{eq:mass_flux}
  j^2 = -\frac{[p]}{[h\tau]} \,,
\end{equation}
with $h$ and $\tau$ being functions of $p$ and $S$. In writing
equation (\ref{eq:zeta}) and in what follows we adopted the convention of
\cite{PMM00} with the plus (minus) sign corresponding to
$S=R$ ($S=L$). Explicit expressions for $j^2(p,S)$ for four different
EoS are given in \S\ref{sec:EoS}.

We now express the post-shock velocity as a function of the post-shock
pressure. We use equations (\ref{eq:zeta}) and (\ref{eq:rankine2}) with
$h\gamma$ in the post-shock state given by equation (\ref{eq:rankine5}) and
$1/D$ given by equation (\ref{eq:rankine1}) \citep{MM96}:
\begin{equation}\label{eq:velocity}
  v(p,S) = \frac{\textstyle h_S\gamma_S v_S + [p]\zeta(p,S)}
		 {\textstyle h_S\gamma_S+[p]\Big(v_S\zeta(p,S) + V_S\Big)}\, .
\end{equation}

\begin{figure}
  \epsscale{0.65}
  \plotone{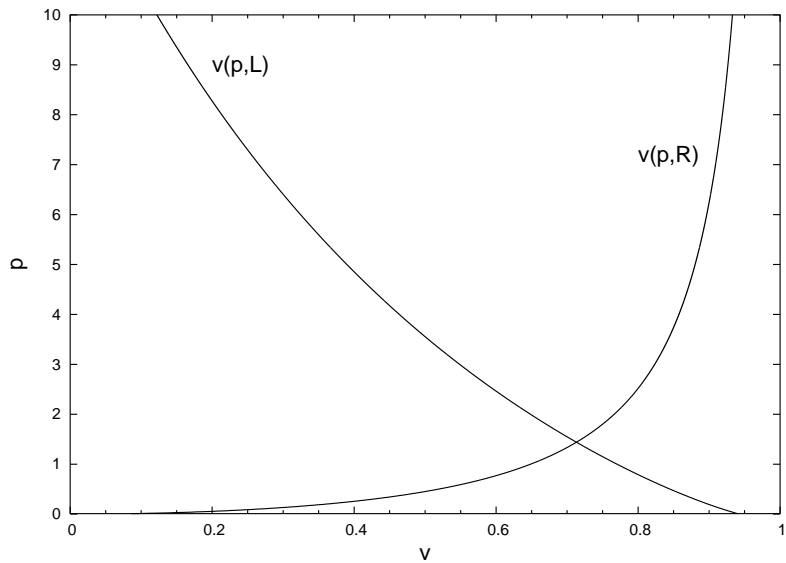}
 \caption{\footnotesize Graphical solution in the $v-p$ plane of $v(p,R)$
           and $v(p,L)$; the intersection between the two curves 
           is the solution to the Riemann problem in the two-shock 
           approximation. \label{fig:riemann_curves}}
\end{figure}

The solution to the Riemann problem is now completely parameterized in
terms of the post-shock pressure $p$. The solution for the normal velocity
and pressure, $v^*$ and $p^*$, has to be continuous across the contact
discontinuity. We achieve this by imposing the condition $v(p,R) = v(p,L)$
as shown in Figure \ref{fig:riemann_curves}.
Since this equation cannot be solved in closed form, an iterative
scheme has to be employed. Here we propose a Newton-Raphson algorithm.
Given the $(n)$-th iteration to $p^*$, the $(n+1)$ approximation is
found through
\begin{equation}\label{eq:newton}
   p^{(n+1)} = p^{(n)} - \frac{v\big(p^{(n)},L\big) - v\big(p^{(n)},R\big)}
                              {v'(p^{(n)},L) - v'(p^{(n)},R)}  \,.
\end{equation}

The explicit expression for the derivative $v'(p^{(n)},S) =
dv(p,S)/dp|_{p=p^{(n)}}$ is found by differentiating equation 
(\ref{eq:velocity}) with respect to the post-shock pressure $p$:
\begin{equation}\label{eq:derivative}
   v'(p,S) \equiv \frac{dv(p,S)}{dp} =
 \frac{\Big(\zeta(p,S) + [p]\zeta'(p,S)\Big)\Big(1 - v_Sv(p,S)\Big) - v(p,S)V_S}
	    {h_S\gamma_S + [p]\Big(\zeta(p,S)v_S + V_S\Big)}  \,.
\end{equation}
Here we take advantage of the fact that the derivative of $\zeta$,
$\zeta'(p,S) \equiv d\zeta(p,S)/dp$, appears only through
$[p]\zeta'(p,S)$. This term can be eliminated by differentiating
equation (\ref{eq:zeta}) with respect to the post-shock pressure and
multiplying the result by $[p]$. With some algebra we obtain
\begin{equation}\label{eq:dp_dzeta}
  [p]\zeta'(p,S) = -\frac{1}{2}\,
         \frac{d(h\tau)/dp + 1/j^2(p,S)}{\zeta(p,S)(1 - v_S^2) - V_S v_S}  \,.
\end{equation}
Analytic expressions for $d(h\tau)/dp$ for selected EoS are given in 
\S\ref{sec:EoS}. When no analytic expression is available or is
expensive to compute, we replace the explicit expression for $v'(p,S)$
with a numerical approximation based on the most recent iteration:
\begin{equation}\label{eq:secant}
   v'(p^{(n)},S) \approx  \frac{v(p^{(n)},S) - v(p^{(n-1)},S)}{p^{(n)} - p^{(n-1)}} \,.
\end{equation}
To start the pressure iteration we use
\begin{equation}
  v'(p_S,S) = \pm\frac{\sqrt{V_S^2 + (1 - v_S^2)/j^2(p_S,S)}}{h_S\gamma_S} \,,
\end{equation}
with the $+$ ($-$) sign for $S = R$ ($S = L$). Note that the term
$j^2(p_S,S)$ appearing in the above equation cannot be computed using
equation (\ref{eq:mass_flux}) since this expression becomes ill-defined when
$p\to p_S$. We address this problem in \S\ref{sec:EoS}.
 
The iteration process stops once the relative change in pressure
falls below $10^{-4}$-$10^{-6}$. Usually no more than 6-7 
iterations are needed to achieve convergence even for the most 
extreme test cases presented in \S\ref{sec:tests}.

Once $v^*$ and $p^*$ have been obtained, the tangential 
velocities $v^*_{2S}$ and $v^*_{3S}$ on both sides of the contact 
discontinuity are computed using equations (\ref{eq:rankine3}) 
and (\ref{eq:rankine4}) with $h\gamma$ in the post-shock state
given by equation (\ref{eq:rankine5}) and
$1/D$ given by equation (\ref{eq:rankine1}):
\begin{equation}\label{eq:tang_vel}
 v^*_{2S,3S} = \frac{\gamma_S h_S v_{2S,3S}}
	      {\gamma_Sh_S + (p^* - p_S)\left(v_{S}\zeta(p^*,S) + V_S\right)}\,.
\end{equation}
Finally we use equation (\ref{eq:rankine1}) to obtain the proper densities
$\rho^*_S$.

The solution to the Riemann problem is now represented by the four
regions in the $(\xi,t)$ plane: $\A{V}_L$, $\A{V}_L^*$, $\A{V}_R^*$
and $\A{V}_R$ (Fig. \ref{fig:riemann_fan}).  The numerical fluxes required
in \S\ref{sec:evolution} are functions of the state vector 
$\Riemann(\A{V}_L, \A{V}_R)$, computed by sampling the solution to
the Riemann problem at the $\xi/t = 0$ axis. To this purpose, 
we define $\sigma = -\sgn(v^*)$; it follows that
\begin{equation}\label{eq:riemann_sample}
 \Riemann\left(\A{V}_L, \A{V}_R\right) = \left\{\begin{array}{cl}
    \A{V}_S^*     &  \qquad \textrm{if}\quad \sigma\lambda_S^* > 0 \,, \\ \noalign{\medskip}
    \A{V}_S       &  \qquad \textrm{if}\quad \sigma\lambda_S   < 0 \,, \\ \noalign{\medskip}
  \DS \frac{\lambda_S\A{V}_S^* - \lambda_S^*\A{V}_S}
           {\lambda_S - \lambda_S^*} 
                  &  \qquad \textrm{if}\quad \sigma\lambda_S^* < 0 < \sigma\lambda_S  \,,
 \end{array}\right. 
\end{equation}
where we set
\begin{equation}
  \A{V}_S^*, \A{V}_S = \left\{\begin{array}{l}
  \A{V}^*_L, \A{V}_L \qquad \textrm{if}\quad \sigma < 0  \,,\\ \noalign{\medskip}
  \A{V}^*_R, \A{V}_R \qquad \textrm{if}\quad \sigma > 0  \,.\\ 
 \end{array}\right.  
\end{equation}

The last condition in equation (\ref{eq:riemann_sample}) holds 
when the solution is sampled inside a rarefaction fan.
In this case we obtain $\Riemann(\A{V}_L, \A{V}_R)$
by linearly interpolating the fan between the head and the tail.

We compute $\lambda_S^*$ and $\lambda_S$ according to whether the wave
separating $\A{V}_S$ from $\A{V}_S^*$ is a shock or a rarefaction.
If $p^* > p_S$ then the wave is a shock, and we set both $\lambda_S$ and
$\lambda_S^*$ to be equal to the shock speed
\begin{equation}
  \lambda_S = \lambda^*_S = \frac{1}{D_S\zeta(p^*,S)} + v_S \,.
\end{equation}
If, on the other hand, $p^* < p_S$ the wave separating 
$\A{V}_S$ from $\A{V}_S^*$ is a rarefaction. 
In this case we first define
\begin{equation}
  \lambda_S^* = \left\{\begin{array}{cc}
    \lambda^-(\A{V}_S^*)  &  \qquad \textrm{if} \quad \sigma < 0 \,,  \\ \noalign{\medskip}
    \lambda^+(\A{V}_R^*)  &  \qquad \textrm{if} \quad \sigma > 0 \,,  \\ 
  \end{array}\right.
\end{equation}
and then enforce the condition $\lambda^-_L < \lambda^*_L$ (if $\sigma < 0$) or 
$\lambda^+_R > \lambda^*_R$ (if $\sigma > 0$) by setting
\begin{equation}
  \lambda_S  = \left\{\begin{array}{cc}
 \min\Big(\lambda^-\left(\A{V}_L\right), \lambda_L^*\Big)
         &  \qquad \textrm{if} \quad \sigma < 0            \,,  \\ \noalign{\medskip}
 \max\Big(\lambda^+\left(\A{V}_R\right), \lambda_R^*\Big)
         &   \qquad \textrm{if} \quad \sigma > 0           \,.  \\ 
  \end{array}\right.
\end{equation}
The expressions for the maximum and minimum wavespeeds $\lambda^+$ 
and $\lambda^-$ are given by equation (\ref{eq:eigenvalues}).

\subsection{Equation of State}\label{sec:EoS}
%
%
%
%
%

The results from the previous section have shown that our Riemann
solver can be written in terms of general expressions involving the
mass flux (eq. [\ref{eq:mass_flux}]), the solution to the Taub adiabat
(eq. [\ref{eq:taub}]), and the derivative $d(h\tau)/dp$
(eq. [\ref{eq:dp_dzeta}]).  Their explicit forms, however, depend on
the particular choice of the EoS. In what follows we consider four
different equations of state that can be cast in the form
$f(h,\Theta)= 0$, where $\Theta = p\tau$ is a temperature-like
variable. EoS properties are conveniently described in terms of the
function 
\begin{equation}\label{eq:gamma_r}
  \Gamma_r(\Theta) \equiv \frac{h(\Theta) - 1}{\Theta} \,.
\end{equation}
Finally, we also derive some useful expressions in constructing
our special relativistic hydrodynamics code.

The first equation of state is represented by the ideal gas EoS
(hereafter labelled $ID$) characterized by a constant polytropic index
$\Gamma$. In this case
\begin{equation}\label{eq:ID}
   h(\Theta) = 1 + \frac{\Gamma}{\Gamma-1} \Theta \,.
\end{equation}

\begin{figure}\centering
 \epsscale{0.65}
 \plotone{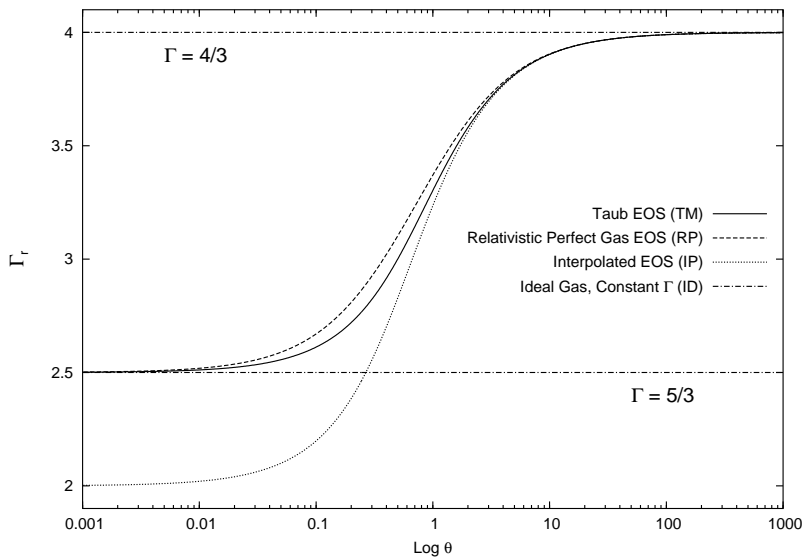}
 \caption{\footnotesize Plots of the function $\Gamma_r(\Theta)$ vs. temperature
              logarithm ($\log\Theta$, $\Theta=p\tau$) for the four different EoS
              discussed in the text.
              The $TM$ EoS (solid line) marks the lower boundary of Taub's inequality
              (\ref{eq:taub_ineq}). The relativistic perfect gas EoS
              ($RP$, eq. [\ref{eq:RP}]) always lies above the boundary
              line, and therefore satisfies Taub's inequality for
              every $\Theta$. The interpolated EoS ($IP$,
              eq. [\ref{eq:IP}]) approaches the correct limit only at high temperatures.
              The horizontal lines correspond to the ideal EoS plotted for $\Gamma=5/3$
              and $\Gamma = 4/3$, but only the second one satisfies Taub's
              inequality for $0\le\theta<\infty$. \label{fig:EoS}}
\end{figure}

For this EoS, $\Gamma_r$ is simply a constant and varies between
$5/2$ (for $\Gamma = 5/3$) and $4$ (for $\Gamma = 4/3$). The ideal gas
EoS is very popular in non-relativistic fluid dynamics and has been
largely used also in studying relativistic flows. However,
equation (\ref{eq:ID}) is not consistent with relativistic formulation of
the kinetic theory of gases and admits superluminal wave propagation
when $\Gamma > 2$ \citep{Taub48}. In his fundamental inequality, Taub proved that the
choice of the EoS for a relativistic gas is not arbitrary. The
admissible region is defined by
\begin{equation}\label{eq:taub_ineq}
  (h - \Theta)(h - 4\Theta) \ge 1 \,,
\end{equation}
shown in the upper-left portion in Figure \ref{fig:EoS}.
Note that for a constant $\Gamma$-law gas, the above condition is
always fulfilled for $\Gamma \le 4/3$ while it cannot be satisfied for
$\Gamma \ge 5/3$ for any $\Theta > 0$.
Furthermore, by inspecting the expressions given in Table \ref{tab:EoS1}
one can see that the ideal gas EoS admits superluminal sound speed 
when $\Gamma > 2$ (corresponding to $\Gamma_r < 2$).
The exact form of relativistic ideal gas EoS was given by \citet{Synge57}.
For a single specie the relativistic perfect ($RP$) gas is described by
\begin{equation}\label{eq:RP}
  h = \frac{K_3(1/\Theta)}{K_2(1/\Theta)}\,,
\end{equation}
where $K_2$ and $K_3$ are the modified Bessel functions of the second
and third kind, respectively.
For the RP EoS $\Gamma_r(\Theta)$ reduces to $5/2$ in the limit of low
temperature ($\Theta\to 0$) while for an extremely hot gas
($\Theta\to\infty$) $\Gamma_r(\Theta)\to 4$.

Unfortunately no simple analytical expression is available for this
EoS and the thermodynamics of the fluid is entirely expressed in terms
of the modified Bessel functions \citep{FK96}.  The
consistency comes at the price of extra computational cost since the
EoS is frequently used in the process of obtaining numerical solution.
Below we offer two more formulations which use analytic expressions
and are therefore more suitable for numerical computations.

\citet{SZS01} proposed a simplified EoS which has the
correct asymptotic behavior in the limit of ultrarelativistic
temperatures ($\Theta\to\infty$).  Their ``interpolated'' EoS ($IP$)
\begin{equation}\label{eq:IP}
  h(h - 4\Theta) = 1 \,,
\end{equation}
has the attractive feature of being simple to implement and is more
realistic than the ideal EoS (eq. [\ref{eq:ID}]). However, the $IP$ EoS
differs from the exact $RP$ EoS in the limit of low temperatures where
$\Gamma_r(\Theta) \to 2$ rather than $\Gamma_r(\Theta) \to
5/2$. Besides, the $IP$ EoS does not satisfy Taub's inequality
(eq. [\ref{eq:taub_ineq}]) as can be seen from Figure (\ref{fig:EoS}).

Here we propose a new EoS which is consistent with Taub's
inequality for all temperatures and it has the correct limiting
values.  Our newly proposed EoS ($TM$) is obtained by taking the equal
sign in equation (\ref{eq:taub_ineq}) to obtain
\begin{equation}\label{eq:TM}
  (h - \Theta)(h - 4\Theta) = 1 \,.
\end{equation}

The $TM$ EoS is quadratic in $h$ and can be solved analytically at
almost no computational cost.
We point out that only the solution with the plus sign 
is physically admissible, since it satisfies $h>1$ for all 
$\Theta$ and it has the right asymptotic 
values for $\Gamma_r(\Theta)$.
Direct evaluation of $\Gamma_r(\Theta)$ shows that the $TM$ 
EoS differs by less than $4\%$
from the theoretical value given by the relativistic perfect gas EoS
(eq. [\ref{eq:RP}]).

For a given EoS, our Riemann solver requires
the solution of the Taub adiabat (eq. [\ref{eq:taub}]), followed
by the calculation of the mass flux (eq. [\ref{eq:mass_flux}]) and
the derivative of $h\tau$ with respect to pressure appearing
in equation (\ref{eq:dp_dzeta}).

For the $ID$ EoS (eq. [\ref{eq:ID}]) and the $TM$ EoS (eq. [\ref{eq:TM}]) the 
Taub adiabat reduces to a quadratic equation in $x=h$ and $x=[h\tau]$,
respectively:
\begin{equation}
    ax^2 + bx + c = 0 \,.
\end{equation}
For the $ID$ EoS the coefficients $a$, $b$ and $c$ are given by
\begin{equation}\label{eq:abc_ID}
   a = 1 - \frac{[p]}{\Gamma_rp}  \,,  \quad
   b = \frac{[p]}{\Gamma_rp}      \,,  \quad
   c = -h_S\big(h_S + \tau_S[p])  \,,
\end{equation}
while for the $TM$ EoS one has
\begin{equation}\label{eq:abc_TM}
 \begin{array}{c} 
  a = p_S(3p + p_S)                     \,,  \quad
  b =  -h^2_S(3p + 2p_S) - h_S\tau_S(3p^2 - 7pp_S - 4p_S^2) - [p] \,,
 \\ \noalign{\medskip} 
    c = - h_S\tau_S[p]\Big(h^2_S  +  2h_S\tau_Sp + 2\Big)\,.
 \end{array}
\end{equation}
The solution is even simpler for the $IP$ EoS (eq. [\ref{eq:IP}]) but has
to be found numerically for the $RP$ EoS (eq. [\ref{eq:RP}]), see Table
\ref{tab:EoS1}.

Straightforward evaluation of the mass flux, equation (\ref{eq:mass_flux}), 
leads to numerical difficulties whenever $[p]
\to 0$ because at the same time also $[h\tau] \to 0$.  For all the
equations of state presented here, with the exception of the $RP$ EoS,
one can find alternative expressions that do not suffer from this
drawback. The procedure is to factor out $[p]$ from $[h\tau]$. The
results, obtained after some algebra, are given in Table
\ref{tab:EoS2}. When such a factorization is not possible,
the limiting value of $j^2$ can be found by noticing that
\begin{equation}
 \frac{1}{j^2(p_S,S)} = -d(h\tau)/dp|_{[p]=0}  \,.
\end{equation}
Since from the Taub adiabat one also has $dh/dp|_{[p]=0} =
\tau_S\,$, we obtain the final result
\begin{equation}
 j^2(p_S,S) \equiv\lim_{[p]\to 0} j^2(p,S)
    = -\frac{1}{\tau_S}\frac{\dot{h}_Sp_S}
                                         {\dot{h}_Sp_S\tau_S + h_S(1 - \dot{h}_S)} \,,
\end{equation}
where $\dot{h}_S = dh(\Theta)/d\Theta |_{[p]=0}$.
Finally, the derivative $d(h\tau)/dp$ is obtained by straightforward
differentiation.

\section{NUMERICAL TESTS}\label{sec:tests}
%
%
%
%
%

We use several problems to verify the correctness of the algorithm and
its implementation. We consider one- and multi-dimensional setups
and problems close to real astrophysical applications.

The first two problems are of particular interest in the context of
Riemann problem as they involve the decay of an initial discontinuity
into three elementary waves: a shock, a contact and a rarefaction
wave. In this case an analytical solution can be found \citep[see][and 
references therein]{PMM00,MM94}. Shock-tube problems are basic 
verification tests useful in assessing the ability of the Riemann
solver to correctly describe evolution of simple waves. 
Thanks to their relative simplicity, one dimensional problems
are also excellent benchmarks that can be used for comparison of
different algorithms and implementations \citep{Calder_etal02}.

We also consider one of the two-dimensional Riemann problems
originally introduced by \citet{SCG93} and later on adopted for relativistic
flows by \citet{dZB02}. The
two-dimensional Riemann test problem is relatively easy to set up and
is characterized by increased complexity. Four elementary waves, two
shocks and two contact discontinuity, are present in the initial
conditions; their mutual interaction is also calculated.

We propose a three-dimensional test involving the reflection of 
a spherical relativistic shock \citep{Aloy99} in cartesian coordinates.

Finally, we consider two astrophysical applications: pressure-matched
light relativistic jet and evolution of the jet in stratified medium.
The latter problem can be relevant in the context of gamma-ray
bursts and their afterglows.

\subsection{Relativistic Shock Tubes}
%
%
%
%
%

In one-dimensional shock-tubes a diaphragm initially separates two
fluids characterized by different hydrodynamic states. At $t=0$ the
diaphragm is removed and gases are allowed to interact. We consider
two cases and assume an ideal equation of state with $\Gamma=5/3$. The
calculations are performed in cartesian geometry. The domain covers
the region $0\leq x\leq 1$ and the discontinuity separating the two
states is initially placed at $x=0.5$. Free outflow is allowed at the
grid boundaries.

\begin{figure}
 \epsscale{0.65}
\plotone{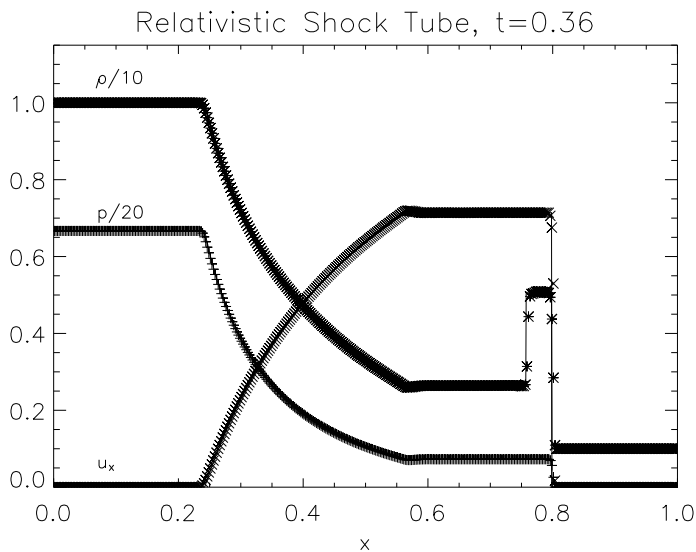}
\caption{\footnotesize The numerical solution (given by the symbols)
         compared to the exact analytical solution (solid line) for
         the relativistic shock tube (\emph{P1} in the text),
         at t=0.36 is shown.
         The wave structure is well described by our numerical scheme:
         both the shock and the contact wave are smeared out on $4-5$
         and $2-3$ zones, respectively, in the density profile
         (asterisk marks). For this test problem $400$ equally spaced
         zones were used and $C_a = 0.9$.\label{fig:shock_tube_1}}
\end{figure}

In the first problem (\emph{P1}) density and pressure to the
left of the interface are $\rho_L = 10$ and $p_L = 40/3$. To the right
of the interface we have $\rho_R = 1$ and $p_R = 2/3\cdot10^{-6}$.
The gas is initially at rest.  This problem has been extensively used
by many authors \citep[see][and reference therein]{MM03};
our results at $t=0.36$ are shown in Figure \ref{fig:shock_tube_1}.

In this problem a rarefaction wave develops and moves to the left while
a contact discontinuity separating the two gases and a shock wave are
both propagating to the right.  The shocked gas accumulates in a thin
shell and is compressed by a factor of $\approx 5$.\footnote{In
relativistic hydrodynamics, the density jump across a shock wave can
attain arbitrary large value.} Relativistic effects are
mainly thermodynamical in nature. The enthalpy of the left state is
appreciably greater than in the non-relativistic limit, $h = 10/3 >
1$, while the Lorentz factor is rather small ($\gamma \approx 1.4$).
With $400$ equidistant zones and $C_a = 0.9$ our algorithm is able
to reproduce the correct wave structure with considerably high
accuracy, as shown in Figure \ref{fig:shock_tube_1}. The shock and the
contact discontinuity are captured correctly. The contact is smeared
only over 2-3 two zones while shock profile occupies 3-4 zones. The
thin dense shell is well resolved. The rarefaction fan is also well
reproduced but a slight overshoot in velocity and undershoot in
density appears at the tail of the rarefaction.
A direct comparison with the analytical solution (see Table 
\ref{tab:L-1error}) shows  
that, at the resolution of $400$ zones, the error measured
in the $L-1$ norm is smaller than $5\%
$. We obtained first order convergence with increasing 
grid resolution, as expected for discontinuous problems.

\begin{figure*}
 \includegraphics[width=5.3cm, height=5cm]{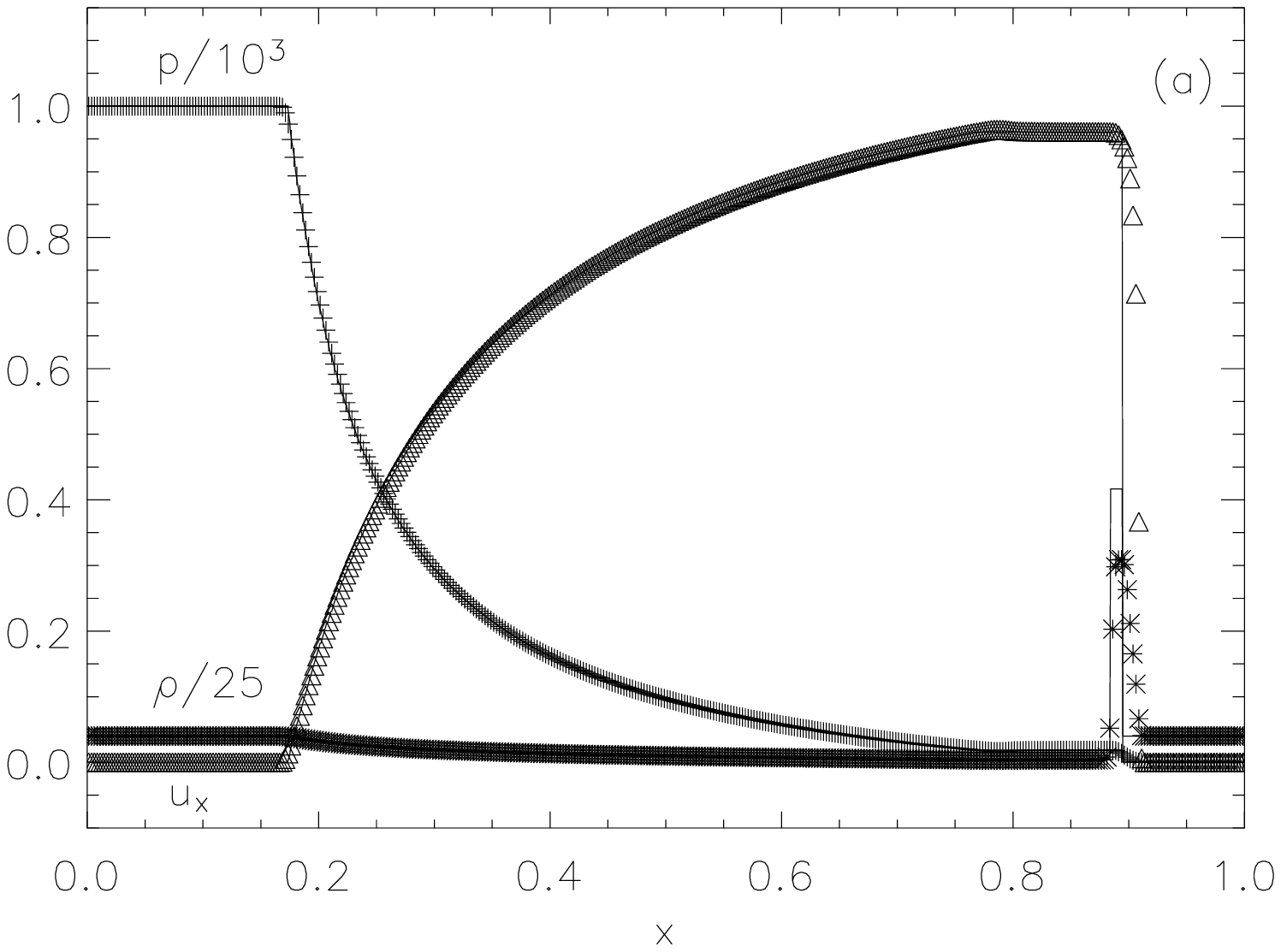}%
 \includegraphics[width=5.3cm, height=5cm]{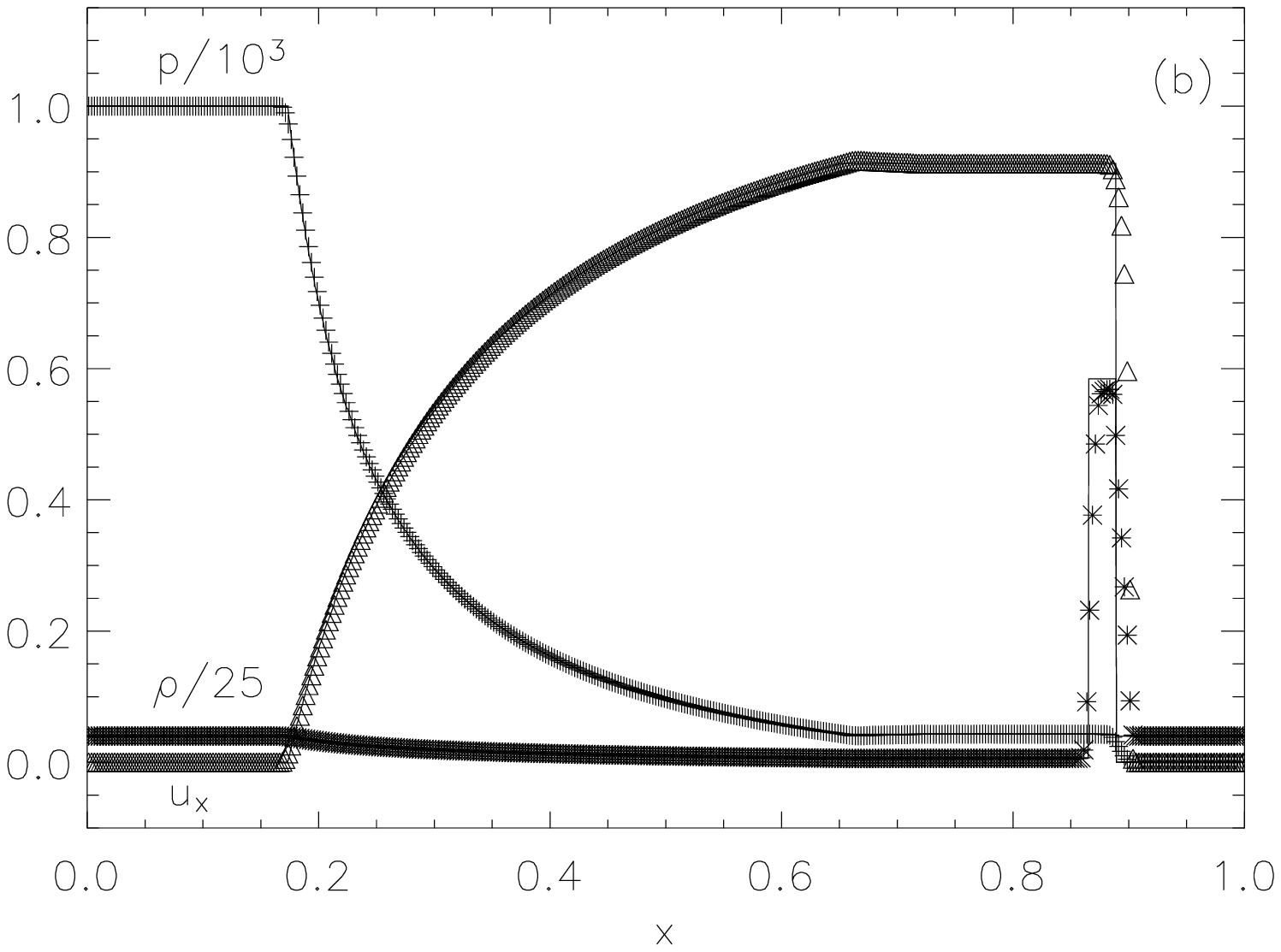}%
 \includegraphics[width=5.3cm, height=5cm]{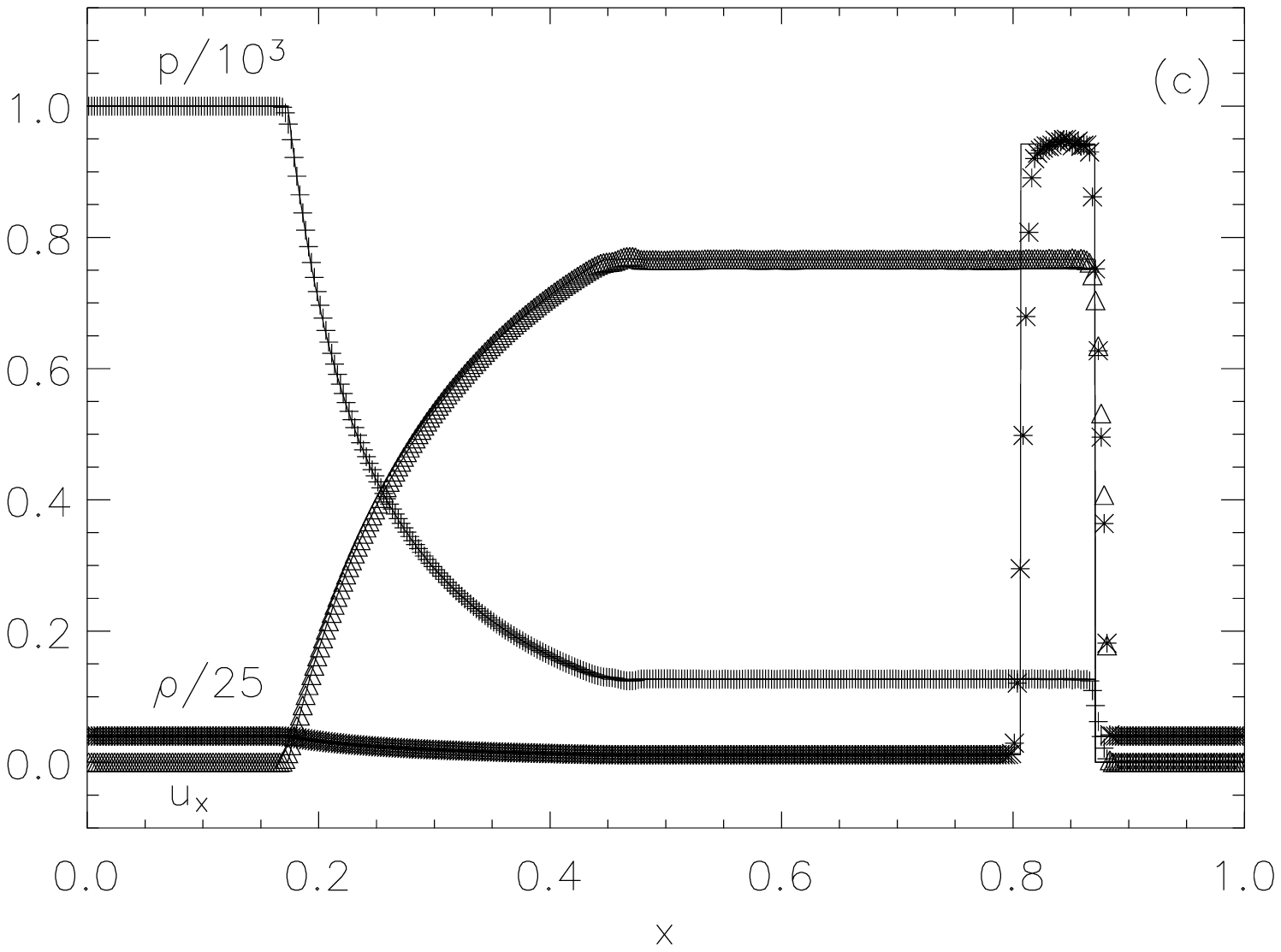}
 \includegraphics[width=5.3cm, height=5cm]{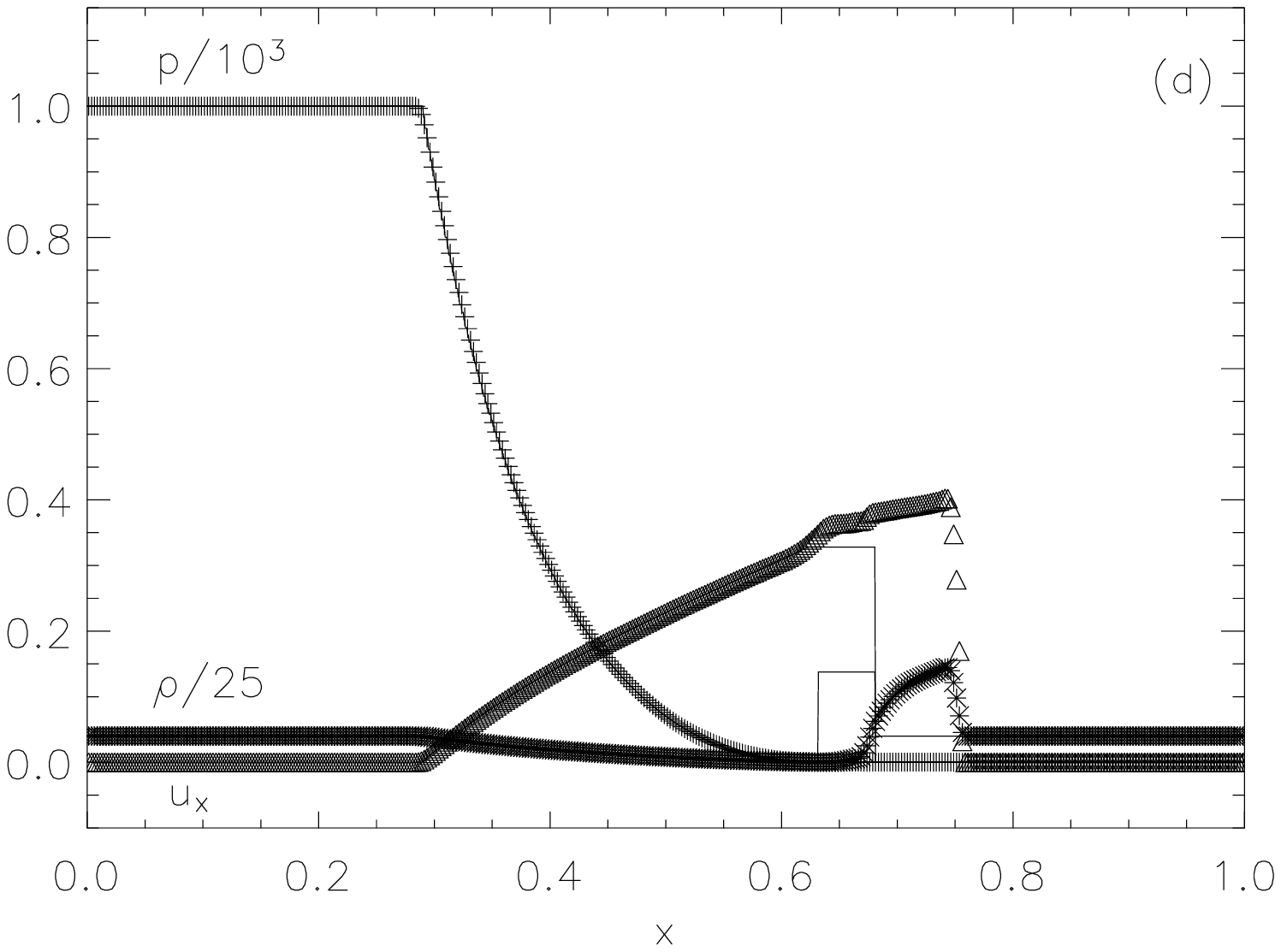}%
 \includegraphics[width=5.3cm, height=5cm]{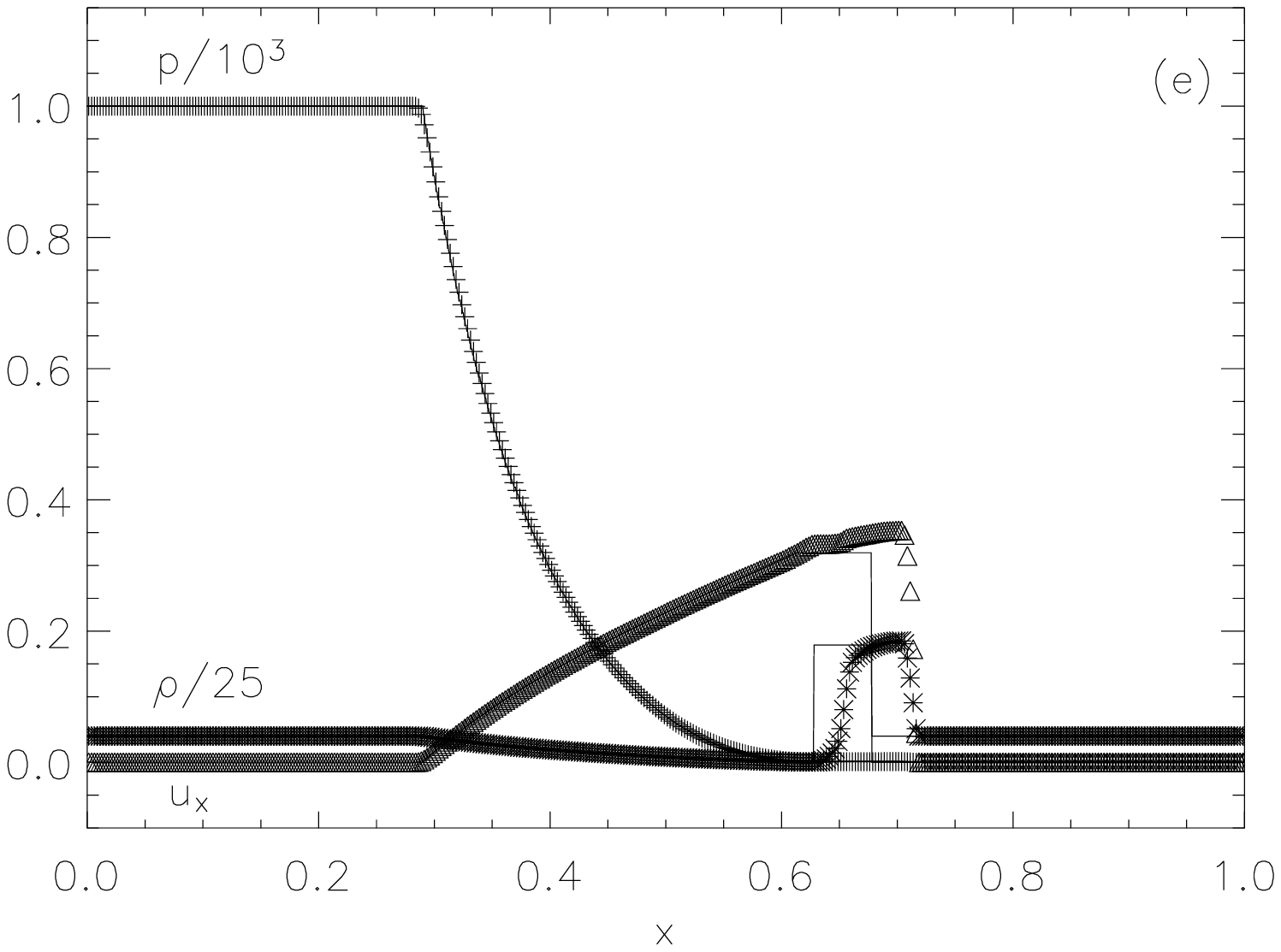}%
 \includegraphics[width=5.3cm, height=5cm]{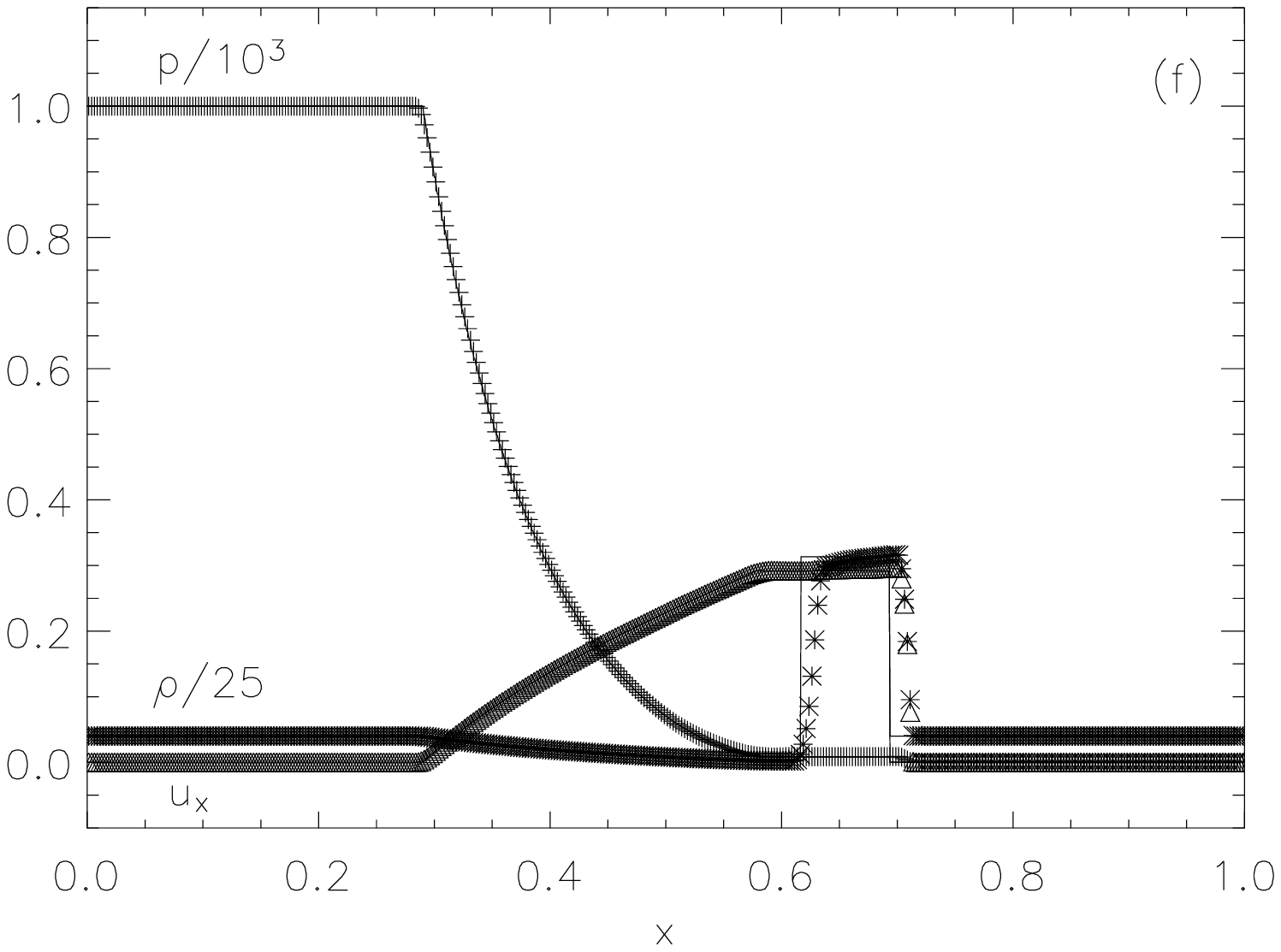}
 \includegraphics[width=5.3cm, height=5cm]{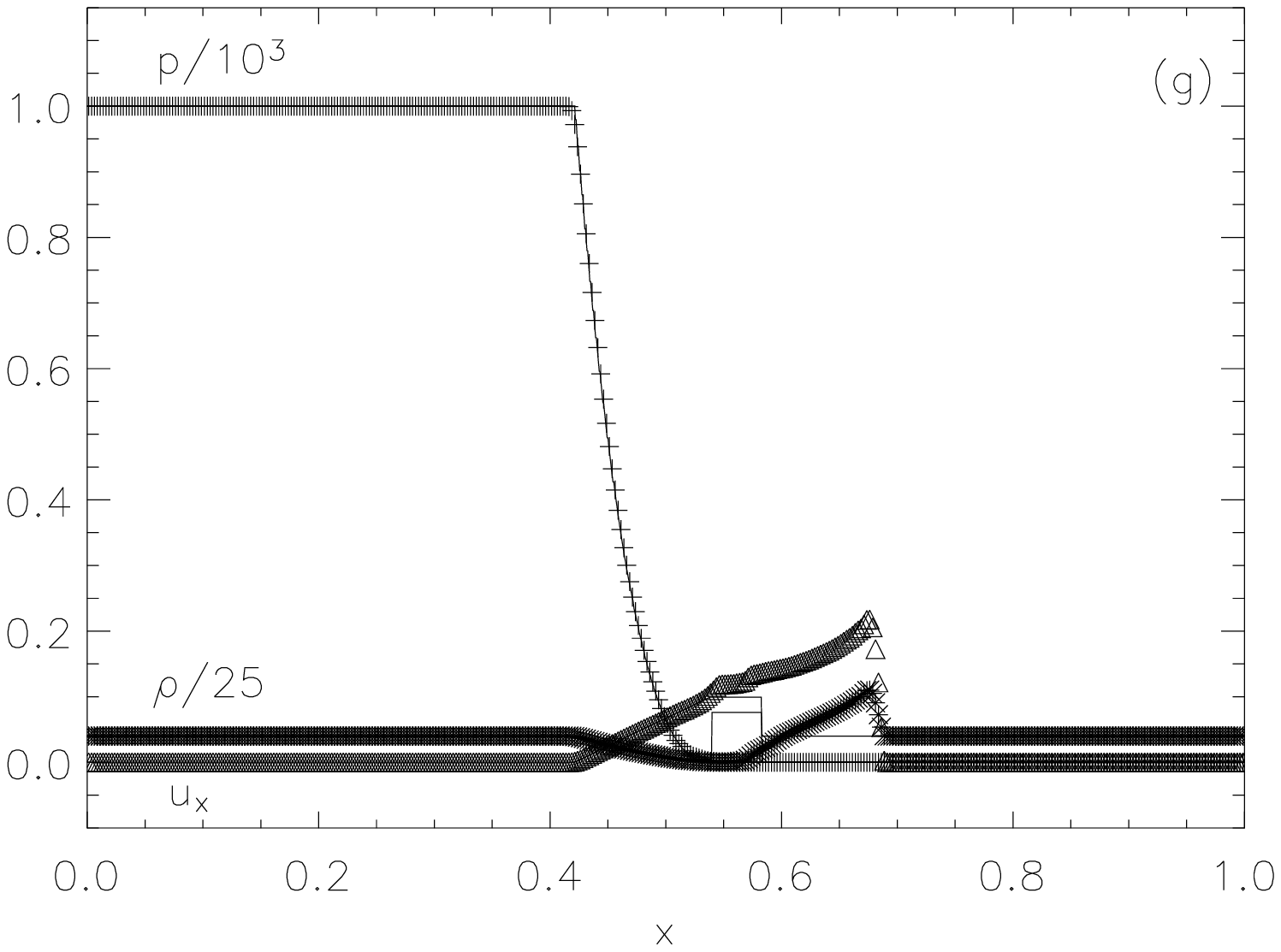}%
 \includegraphics[width=5.3cm, height=5cm]{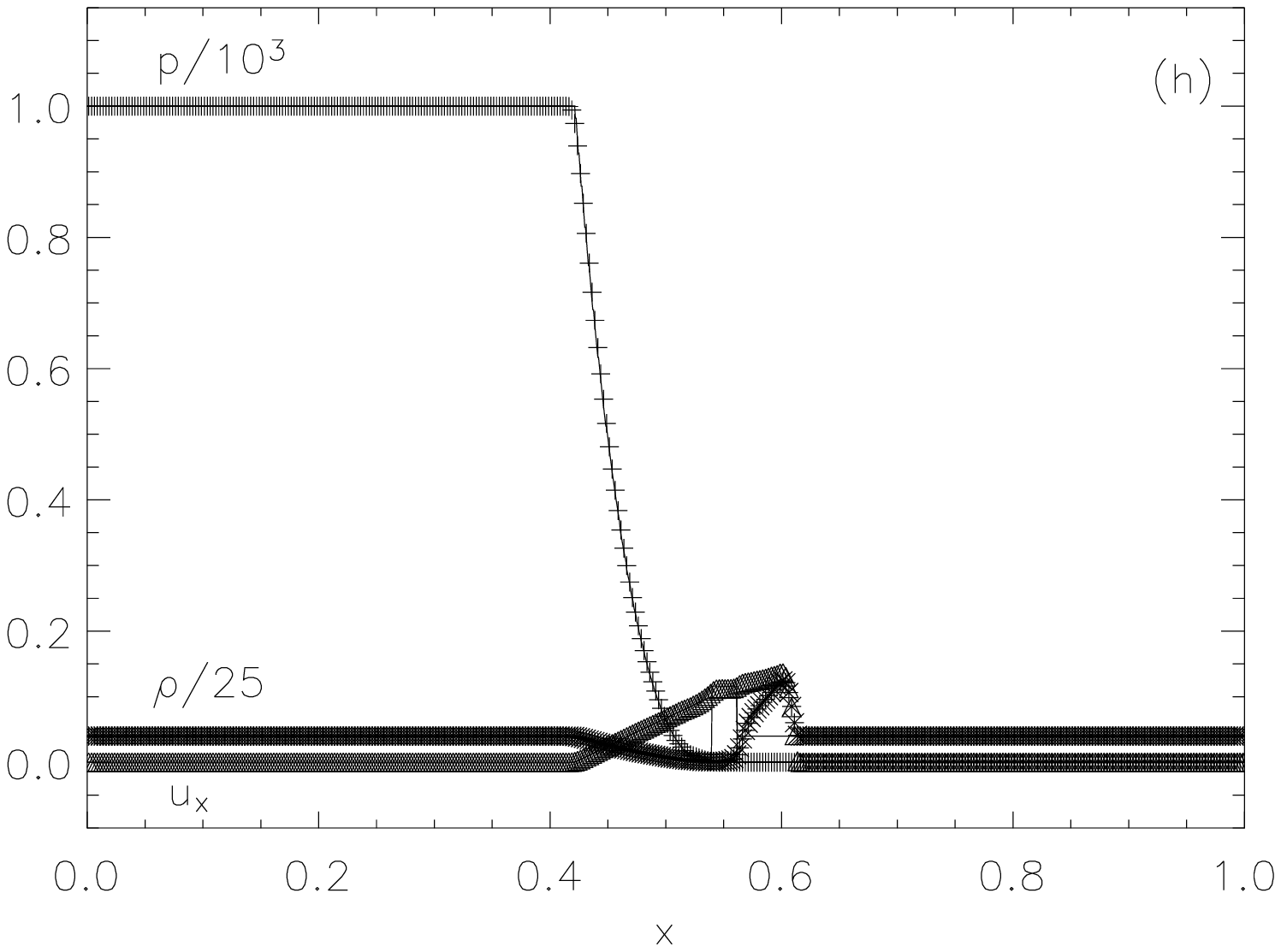}%
 \includegraphics[width=5.3cm, height=5cm]{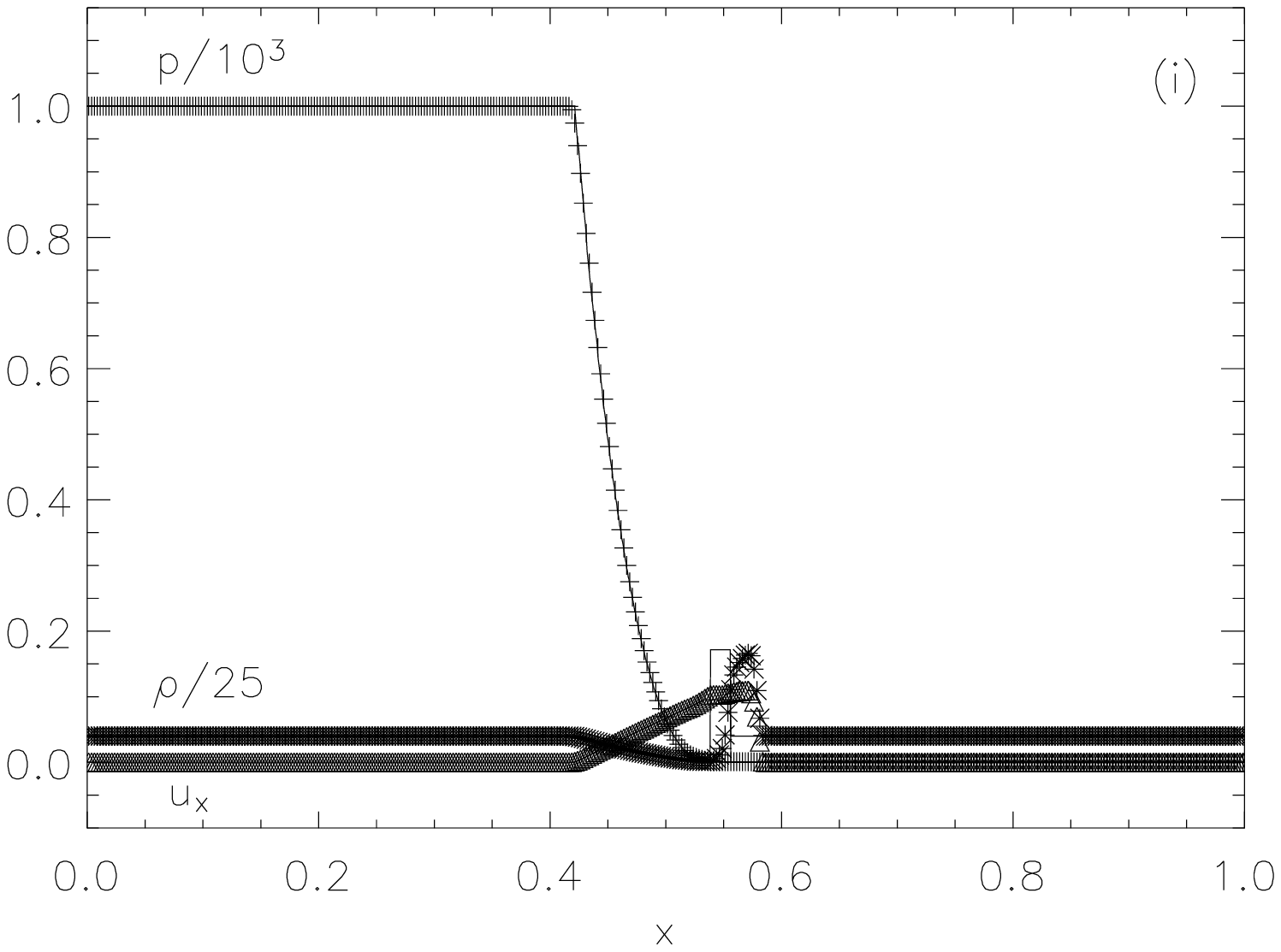}
 \caption{\footnotesize Relativistic shock tube (\emph{P2}): 
          analytical solution (solid lines) and numerical solutions on 
          400 zones are shown at $t=0.4$; from left to right $v_{yR} = 
          0,0.9,0.99$, from top to bottom $v_{yL} = 0,0.9,0.99$.
          \label{fig:shock_tube_2}}
\end{figure*}

In the second one-dimensional test (\emph{P2}) the gas to the left of
the discontinuity has pressure $p_L = 10^3$ while $p_R = 10^{-2}$. The
proper gas density and velocity are uniform everywhere and equal to
$\rho=1$ and $v_x=0$, respectively. The initial conditions are
therefore similar to those of \emph{P1}, but the internal energy of
the left state is greater than the rest-mass energy, $h\approx
2.5\cdot 10^3 \gg 1$, resulting in a thermodynamically relativistic
configuration.  We also include the effects of non vanishing
tangential velocity. The test is run for nine possible combinations of
pairs $v_{y,L}$ and $v_{y,R}$ constructed from the set of $\{0, 0.9,
0.99\}$ \citep{PMM00}. The results at $t = 0.4$ are shown in Figure
\ref{fig:shock_tube_2}.

The decay of the initial discontinuity in \emph{P2} leads to the same
wave pattern as in \emph{P1}, but the changing tangential velocity has
a strong impact on the solution. In absence of shear velocities
(Fig. \ref{fig:shock_tube_2}$a$), the high density shell is extremely
thin posing a serious challenge for any numerical scheme. With $400$
zones and $C_a=0.4$ the post-shock density is underestimated by $\sim
25\%$. The error is reduced to $\sim 6\%$ when the resolution is
doubled.

The solution changes drastically when tangential velocities are
included due to the increased steepness of the shock adiabats, $dp/dv$. This
can be seen from equations (\ref{eq:velocity}) and (\ref{eq:derivative})
setting $v_S = 0$. This is a purely relativistic effect resulting from
the coupling of the specific enthalpy (which is independent of
transverse velocities) and the Lorentz factor in the upstream region.
Specifically, a higher tangential velocity in the right state has the
effect of increasing the post-shock pressure, so that, even if the
effective inertia of the right state is larger (due to the higher
Lorentz factor), the two effects compensate and the shock speed is
only weakly dependent on $v_{yR}$.  As a net result the mass flux
increases with the Lorentz factor of the right state $\gamma_R$. This
leads to the formation of a denser and thicker shell (cases $b$, $c$,
and to smaller degree $e$ and $f$ in Fig. \ref{fig:shock_tube_2}).

The opposite behavior is displayed when the tangential velocity is large
in the left state (cases $d$ and $g$), leading to a lower post-shock
pressure and smaller normal velocity. As a consequence, the shock
strength and the mass flux through the shock are largely reduced.

Figure \ref{fig:shock_tube_2} presents the nine numerical tests
together with the corresponding exact analytical solutions
\citep{PMM00}. One can clearly notice an excessive smearing of the
contact discontinuity with increasing $v_{yL}$. The situation
does not improve even when the exact Riemann solver is used and the
evolution of rarefaction waves is calculated with greater accuracy. We
speculate that the reason for excessive smearing of the contact
discontinuity is that our numerical scheme is not able to properly
resolve waves right after breakup of the initial discontinuity. We
expect that other currently existing shock-capturing schemes
experience similar diffuculties in this case, but such a detailed
study is beyond the scope of the present publication.

\subsection{Two-dimensional Riemann Problem}\label{sec:riemann_2d}
%
%
%
%
%

Two-dimensional Riemann problems have been originally proposed by
\citet{SCG93} in the context of classical
hydrodynamics \citep{LL98}. This set of problems
involves interactions of four elementary waves (shocks, rarefactions,
and contact discontinuities) initially separating four constant states.
A relativistic extension of a configuration involving two shocks and
two contact discontinuities was recently presented by
\citet{dZB02}.  Here we consider this test but with
slightly modified initial conditions that reproduce a simple wave
structure (the conditions adopted by \citet{dZB02} do
not yield elementary waves at every interface).

\begin{figure}
 \epsscale{1.0}
 \plottwo{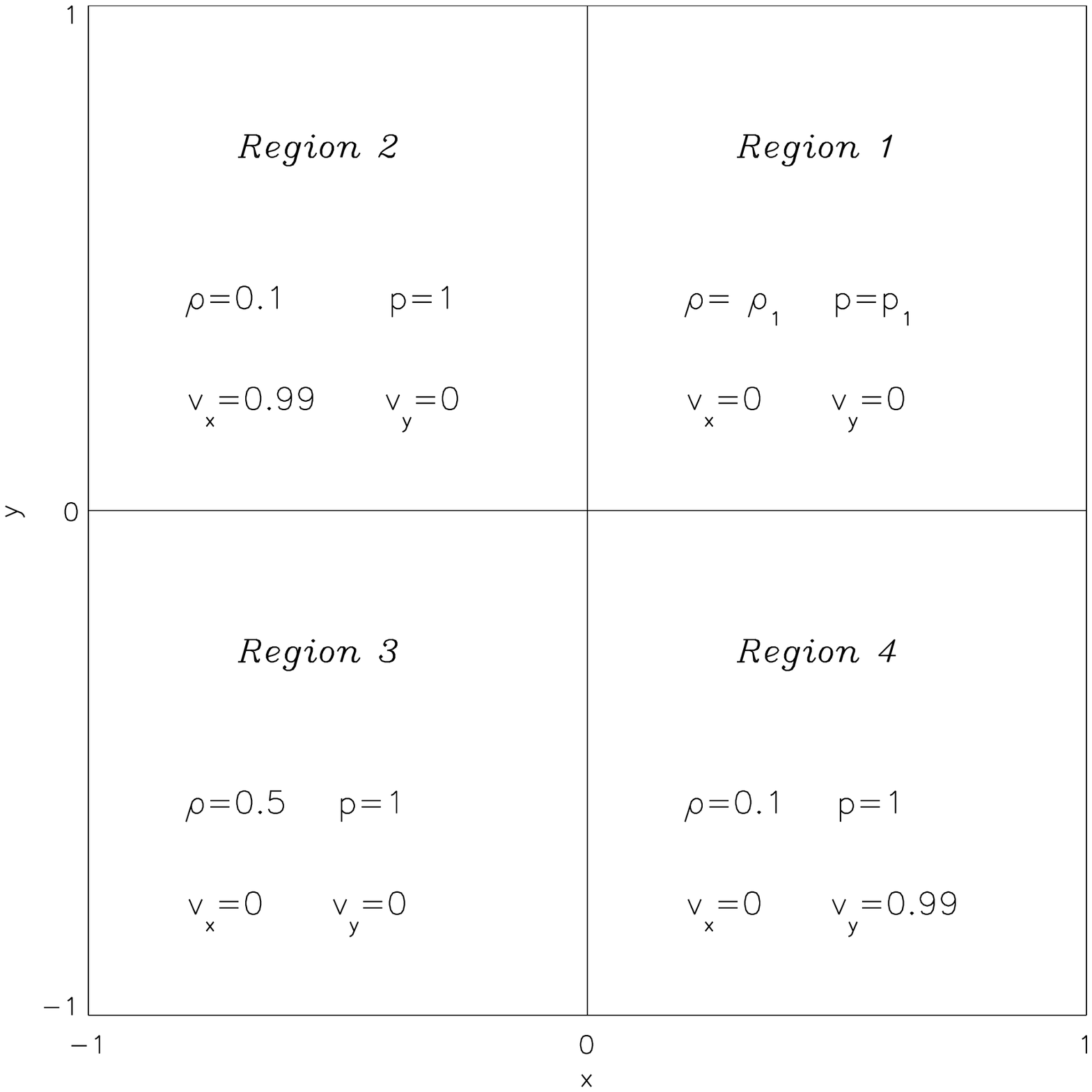}{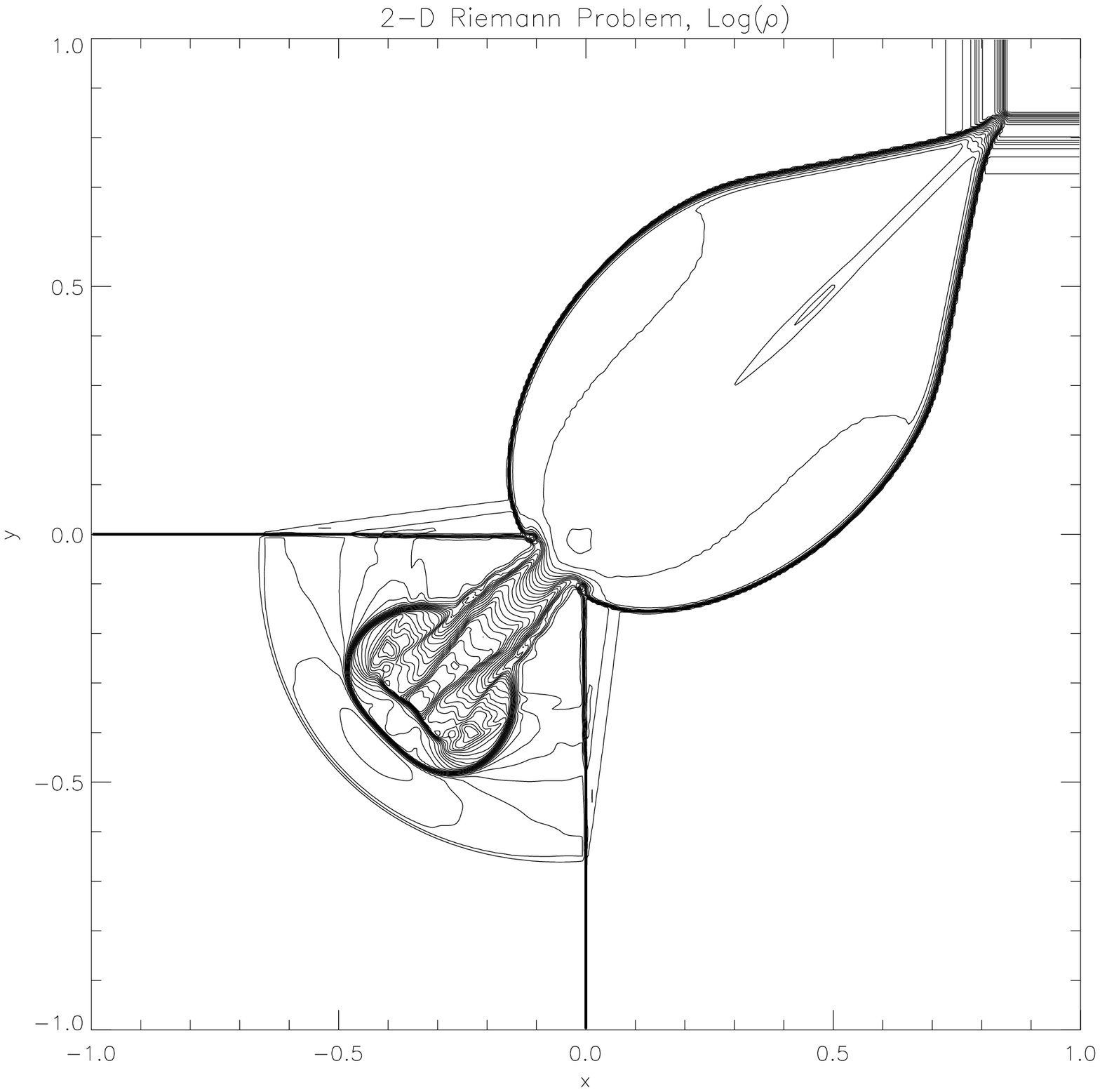}
 \caption{\footnotesize Initial condition for the 2-D riemann problem 
          (on the left) and solution at t= 0.4 (right). In order to
           have simple shocks at the $(2\leftrightarrow1)$, $(4\leftrightarrow1)$
           interfaces, the value of density and pressure in region $1$ are 
           $\rho_1 = 5.477875\cdot 10^{-3}$,
           $p_1 = 2.762987\cdot 10^{-3}$.\label{fig:Riemann2D}}
\end{figure}

We define four constant states in the four quadrants: $x,y >0$ (region
$1$), $x<0, y>0$ (region $2$), $x,y<0$ (region $3$), and $x>0, y<0$
(region $4$). The hydrodynamic state for
each region is given in the left panel in Figure \ref{fig:Riemann2D}
with $\rho_1 = 5.477875\cdot 10^{-3}$ and $p_1 = 2.762987\cdot
10^{-3}$ The computational domain is the box $[-1,1]\times[-1,1]$.  We
used an ideal EoS with $\Gamma = 5/3$. The integration is carried out
with $C_a=0.4$ till $t=0.8$.

The initial conditions result in two equal-strength shock waves
propagating into region $1$ from regions $2$ and $4$ 
(right panel in Fig. \ref{fig:Riemann2D}). The two contact
discontinuities bound region $3$. By $t=0.8$ the unperturbed solution
is still present in part of region $1$ and $3$. Most of the grid
interior is filled with the shocked gas. The two shocks continuously
collide in region $1$. Curved transmitted shock fronts bound a
double-shocked drop-like shaped region located along the main
diagonal. At the final time the symmetry of the problem is well
preserved with a relative accuracy of about $10^{-9}$ in density.  
We also note that our nonlinear Riemann solver 
can capture stationary contact discontinuities exactly, without 
producing the typical smearing of more diffusive algorithms \citep{dZB02}.

\subsection{Relativistic Spherical Shock Reflection}\label{sec:rssr}
%
%
%
%
%

The initial configuration for this test problem consists in a cold ($p = 0$),
uniform ($\rho = 1$) medium with constant spherical inflow velocity 
$v_{in}$ and Lorentz factor $\gamma_{in}$. 
At $t=0$ the gas collides at the center of the computational domain and 
a strong shock wave is formed. For $t>0$ the shock propagates upstream and
the solution has an analytic form given by \citep{MMFIM97}:
\begin{equation}
  \rho(r,t) = \left\{ \begin{array}{cc}
\DS   \left(1 + |v_{in}|\frac{t}{r}\right)^\alpha        &  \; \textrm{for}\quad r > v_st \\  \noalign{\medskip}
\DS   \left(1 + \frac{|v_{in}|}{v_s}\right)^\alpha\sigma &  \; \textrm{for}\quad r < v_st 
  \end{array}\right.
\end{equation}
where 
\begin{equation}
  \sigma = \frac{\Gamma + 1}{\Gamma - 1} + \frac{\Gamma}{\Gamma - 1}\left(\gamma_{in} - 1\right)
\end{equation}
is the compression ratio and 
\begin{equation}
  v_s = \frac{\Gamma - 1}{\gamma + 1} \gamma_{in} |v_{in}|
\end{equation}
is the shock velocity.
Here $\alpha = 0,1,2$ for cartesian, cylindrical and spherical geometry, respectively.
Behind the shock wave ($r < v_st$), the gas is at rest (i.e. $v=0$) and the 
pressure has the constant value $\rho(r,t)(\gamma_{in} - 1)(\Gamma - 1)$.
Conversely, in front of the shock all of the energy is kinetic and thus $p=0$,  
$v = v_{in}$.

This test problem has been proposed by \cite{Aloy99} in three-dimensional
cartesian coordinates to test the ability of the algorithm in keeping  
the spherically symmetric character of the solution.
For numerical reasons, pressure has been initialized to a small finite 
value, $p = 2.29\cdot10^{-5}(\Gamma - 1)$, with $\Gamma = 4/3$.
The computational domain is the cube $[-1,1]^3$ and the final integration
time is $t = 2$.

\begin{figure*}
 \includegraphics[width=0.5\textwidth]{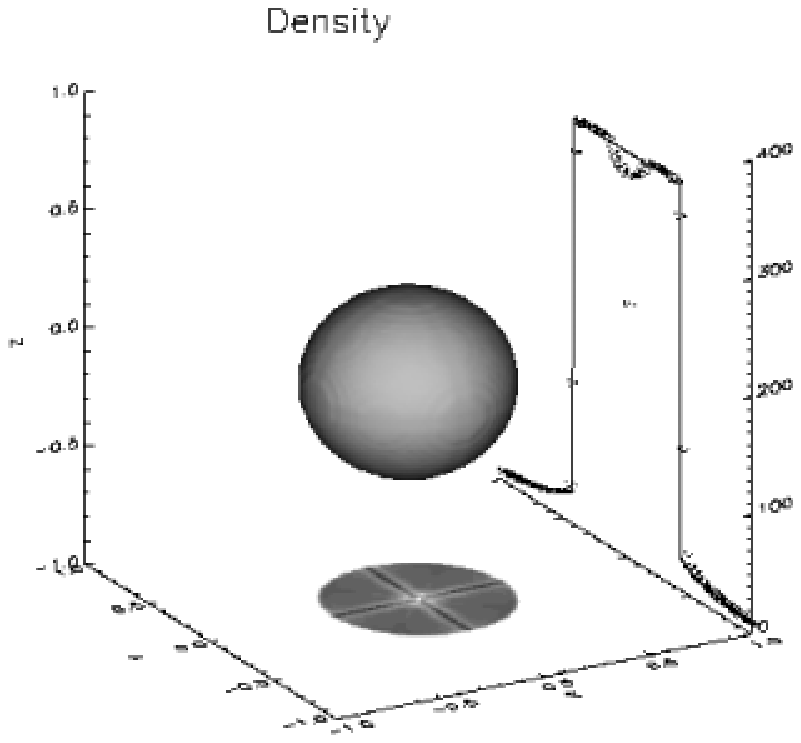}%
 \includegraphics[width=0.5\textwidth]{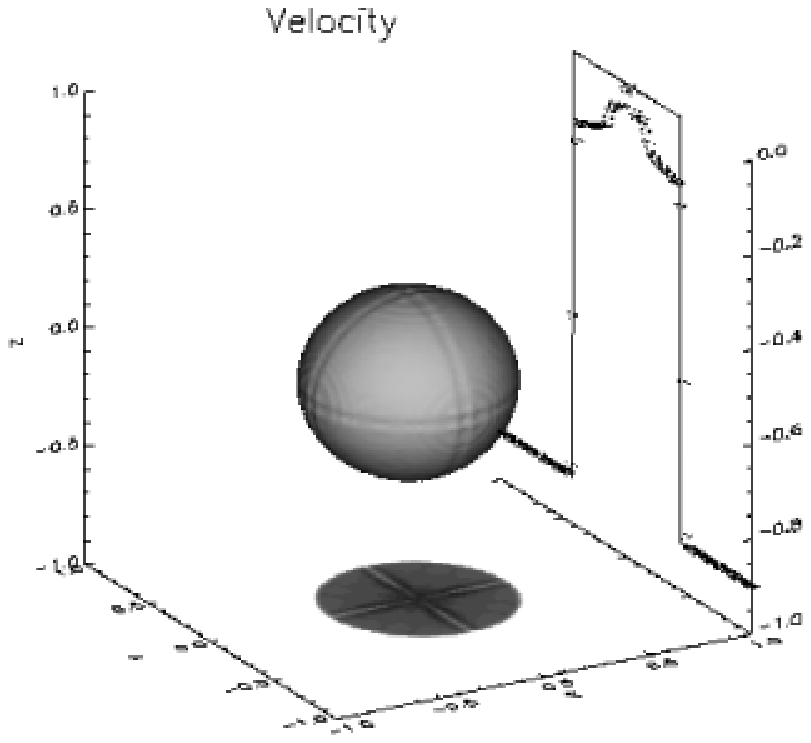}
 \includegraphics[width=0.5\textwidth]{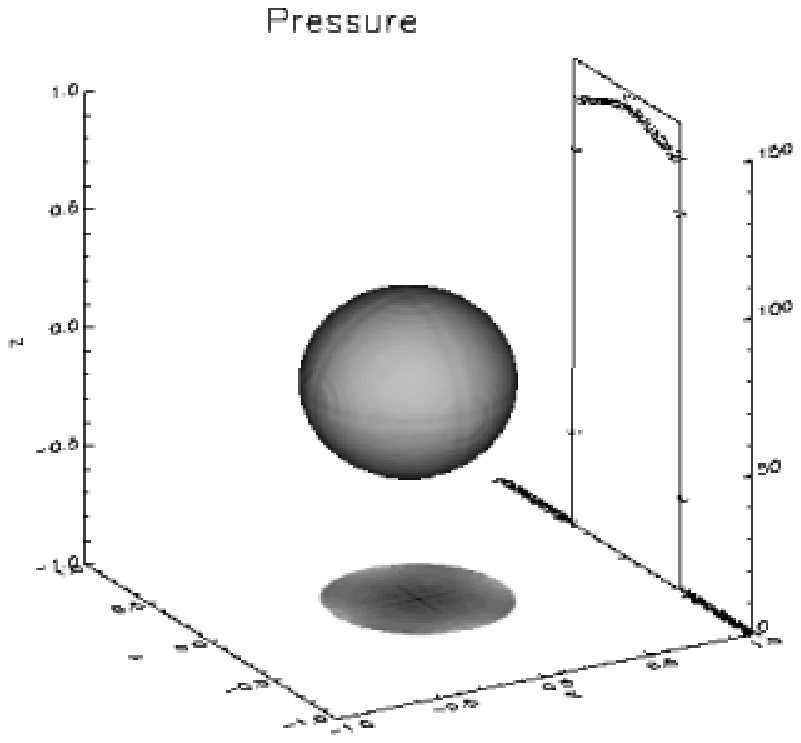}%
 \includegraphics[width=0.5\textwidth]{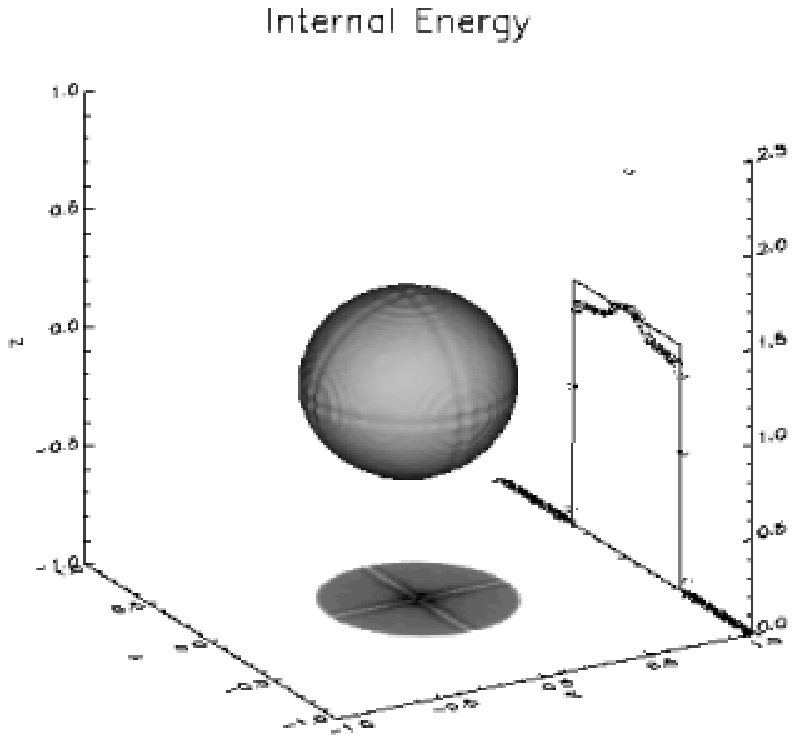}
 \caption{\footnotesize Three dimensional isosurfaces of density
          (top left), total velocity (top right), pressure (bottom
          left) and specific internal energy (bottom right) for
          the relativistic spherical shock reflection test problem 
          (RSSR) at $t=2$. The inflow velocity for this case 
          is $v_{in} = -0.9$.
          The intensity plots below each rendered surface
          show two-dimensional slices of the same quantity in the 
          XY plane at at $z=0$.
          Finally, 
          the numerical (diamonds) and analytic (solid line) 
          solutions at $y,z=0$ are compared in the one dimensional 
          plots projected on the YZ plane. 
          The resolution adopted for this case is $101^3$ and 
          integration has been carried with $CFL = 0.4$.
          \label{fig:rssr}}
\end{figure*}

Figure \ref{fig:rssr} shows the results for $v_{in} = -0.9$ on 
$101^3$ computational zones; the Courant number is $C_a = 0.4$.
The relative global errors \citep[computed as in][]{Aloy99} 
for density, velocity and pressure are, respectively, 
$10.7\%$, $0.7\%$, $16.4\%$. 
Numerical symmetries (computed as $\max(\rho(x_i, y_j, 0) -
 \rho(0, y_k, z_j)$) are retained to $10^{-7}$ for this case. 
From the same Figure we notice that our solver suffers from the 
wall-heating phenomenon, typical of such schemes.  

We also investigated the ultra-relativistic regime by increasing the  
inflow velocity according to $v_{in} = \nu - 1$, where 
$\nu = 10^{-1}, 10^{-3}, 10^{-5},10^{-7}$, the latter corresponding
to a Lorentz factor of $\gamma_{in} \approx 2236$. Following
\cite{Aloy99} we perform this set of simulations on $81^3$ zones.
The relative global errors are given in Table \ref{tab:rssr}.


\subsection{Multi-dimensional Applications}
%
%
%
%
%

Relativistic hydrodynamics is expected to play an important role in
many phenomena typical in high energy astrophysics. In this Section
we present the results of application of our new algorithm to the
problem of propagation of relativistic jets. In the first problem we
consider a highly supersonic pressure-matched light jet propagating in
a constant density medium. In our second example we study the
evolution of a light, high-Lorentz factor jet in a stratified medium,
a situation frequently considered in the context of cosmological
gamma-ray bursts.

\subsubsection{Axisymmetric Jet}
%
%
%
%
%

We consider the propagation of an axisymmetric relativistic jet in 2-D
cylindrical coordinates. The computational domain covers the region $0< r <
15$ and $0< z < 45$ jet radii.  At $t=0$ the jet is placed on the grid
inside the region $z\le 1$, $r\le 1$ with velocity $v_z = 0.995$ parallel to
the symmetry axis and jet beam density $\rho_b = 1$. The ambient medium
has density $\rho_m = 100$ and pressure is uniform everywhere,
$p = 0.005$.

We use the $TM$ equation of state, equation (\ref{eq:TM}), which for 
the current conditions yields a relativistic beam Mach number of 
$\gamma v_z / (c_s\gamma_{c_s}) \approx 110.5$ (corresponding to a 
classical Mach number $v_b/c_{s}\approx 11$).
The computational domain is covered by $360\times 960$
equally spaced zones in the $\nv_r$ and $\nv_z$ direction
respectively. At the symmetry axis we apply reflecting boundary 
conditions. Free outflow is allowed for at all other boundaries 
with the exception of the jet inlet ($z=0$, $r<1$) where $(\rho,\A{v},p)$
are kept at their initial values. The integration is performed
with $C_{a}=0.4$.  

\begin{figure}
  \epsscale{0.5}
  \plotone{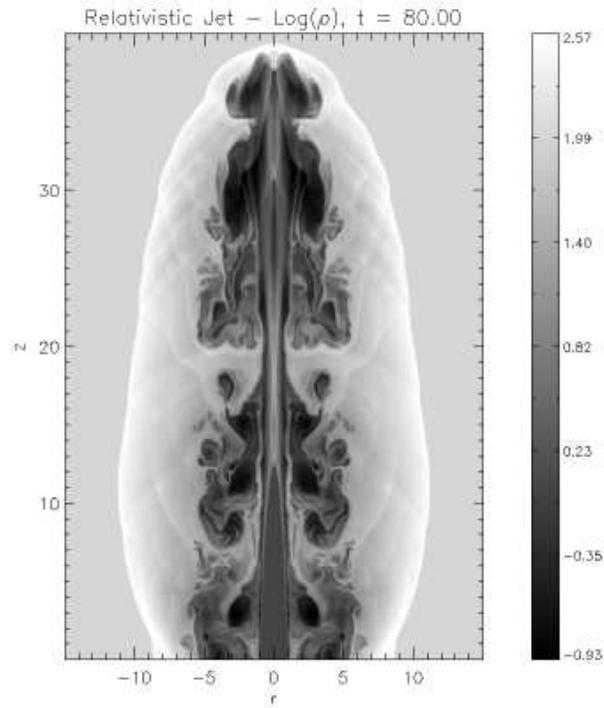}
  \caption{\footnotesize Density distribution, in logarithmic scale, 
           for the axisymmetric relativistic jet at $t=80$. The jet 
           is initially in pressure equilibrium with the ambient 
           medium, which is $100$ times denser than the jet; the 
           initial velocity is $v_z = 0.995$. The resolution is 24
           zones per beam radius.\label{fig:rjet}}
\end{figure}

We follow the evolution until $t = 80$ (in units of jet radius  
crossing time at the speed of light). By that time the jet has
developed a rich and complex structure (Fig. \ref{fig:rjet}). The
morphology of the jet cocoon is determined by the interaction of the
supersonic collimated beam (inside $r\lesssim 1$) with the shocked ambient
gas.  The beam is decelerated by the Mach disk and is terminated by
the contact discontinuity where the beam material is deflected
sideways feeding the cocoon. The interface between the jet backflowing
material and the
ambient shocked gas is Kelvin-Helmholtz unstable resulting in the
formation of vortices and mixing. For the current conditions the jet
cocoon shows a mix of turbulent regions close to the beam and
relatively smooth outer sections traversed by a number of weak sound
waves. The beam shows 3-4 internal shocks in its middle section and
another strong internal shock about 10 jet radii behind the jet head.

\begin{figure}
  \epsscale{0.6}
  \plotone{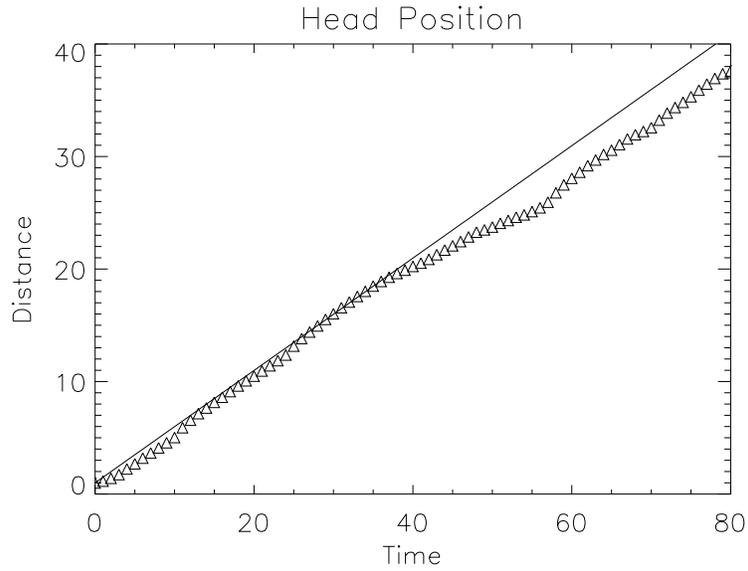}
  \caption{\footnotesize Distance traveled by the jet head
           as a function of time. The triangles show the 
           position of the head in our simulation at different 
           times, while the solid line represents 
           the relativistic one-dimensional estimate.\label{fig:rjet_head_pos}}
\end{figure}

Our results favorably compare to the previous study of
\citet{MMFIM97}.  Although in our simulation we do not use an ideal
EoS, the equivalent adiabatic index, defined through equation (\ref{eq:ID})
as $\Gamma_{eq} = (h-1)/(h-1-\Theta)$, always remains very close to
$5/3$ making direct comparison with other studies
possible. Specifically, our choice of parameters places the presented
model between models $C2$ and $C3$ of \citet{MMFIM97}.
Figure \ref{fig:rjet_head_pos} shows the position of the jet head as
a function of time.  We find that during the early evolution ($t \lesssim
40$) the jet head velocity agrees very well with the one-dimensional
theoretical estimate given by \cite{MMFIM97}.  At later
times, $40\lesssim t \lesssim 60$, the jet enters a deceleration
stage after which ($t \gtrsim 60$) the velocity becomes again
comparable with the theoretical prediction.

The propagation efficiency, defined as the ratio between the average
head velocity and its one-dimensional estimate, is found to be
$\approx 0.92$. This result is consistent with the previous results
obtained for highly-supersonic light jets with large adiabatic index
for which beam velocities tend to have propagation efficiency close to
1 \citep[see Table 2 in][]{MMFIM97}.

\subsubsection{Gamma Ray Burst}
%
%
%
%
%

We propose a simplified model for the propagation of a relativistic
jet through a collapsing, non-rotating massive star. Such objects are
believed to be sources of long-duration gamma ray bursts observed at
cosmological distances \citep{Frail01,WmF99,Aloy00}. 
Our choice of parameters closely reflects that of model JB of 
\citet{ZWmF03}.

We adopt spherical coordinates $(r,\theta)$. The domain has $320$
uniformly spaced zones between $1 \le r/r_0 \le 11$ and $500$
geometrically stretched zones in the range $11 \le r/r_0 \le 500$.
Here $r_0 = 2\cdot 10^8$ cm is the inner grid boundary.  The grid in
the $\theta$ direction consists of $180$ uniformly-spaced zones for
$0^\circ \le \theta\le 30^\circ$, while it is geometrically stretched
with $100$ zones covering the region $30^\circ \le \theta \le
90^\circ$.

For the sake of simplicity, density and pressure distributions in the
stellar progenitor are given by single power-laws,
\begin{equation}\label{eq:GRB_medium}
   \rho_m = \rho_0 \left(\frac{r_0}{r}\right)^a \,,\quad 
   p_m = p_0 \left(\frac{r_0}{r}\right)^b \,,
\end{equation}
while the velocity is zero everywhere.
We adopted $a = 1.77$, $b = 2.6$, $\rho_0 = 10^6$ gr/cm$^3$, and $p_0
= 8.33 \cdot 10^{23}$ dyn/cm$^3$ to match the progenitor model of
\cite{ZWmF03}.
To account for gravity we included an external force 
corresponding to a $5 M_\odot$ gravitational point mass placed 
at $r = 0$. This is accounted for during the source step of
our algorithm.

The jet is injected at $r = r_0$, $0\le \theta \le \theta_0$ where
$\theta_0$ is the jet half-opening angle. The jet is characterized
by its Lorentz factor $\gamma_b$, energy deposition rate (jet power)
\begin{equation}\label{eq:GRB_power}
  \dot{E}_b = \rho_b\gamma_b(h_b\gamma_b - 1)v_b2\pi r_0^2(1 - \cos\theta_0) \,,
\end{equation}
and kinetic to total energy density ratio
\begin{equation}\label{eq:GRB_ratio}
  f = \frac{ \rho_b\gamma_b(\gamma_b - 1)}{\rho_b\gamma_b(h_b\gamma_b - 1) - p_b} \,,
\end{equation}
where $\rho_b$, $h_b$ and $p_b$ are the proper density, 
specific enthalpy and pressure of the beam, respectively. 
Notice that the total energy density considered here does not 
include the rest mass energy.
Equations (\ref{eq:GRB_power}) and (\ref{eq:GRB_ratio}) are used to
express pressure and density as functions of the jet power, fractional
kinetic energy density, Lorentz factor, and jet half-opening
angle.  For the present application we use $\dot{E}_b = 10^{51}$ 
erg s$^{-1}$, $f = 0.33$, $\gamma_b = 50$ and $\theta_0 = 5^\circ$.
The $TM$ equation of state is adopted and the CFL number is $C_a = 0.6$.
Reflecting boundary conditions are used at
$\theta = 0$ and $\theta = 90^\circ$, while we allow for a free
outflow at the outer grid boundary ($r/r_0 = 500$) and at the inner
boundary outside the injection region.  We follow the evolution
until the jet has reached the outer boundary at $r = 10^{11}$ cm.

\begin{figure} 
 \epsscale{1}
 \plottwo{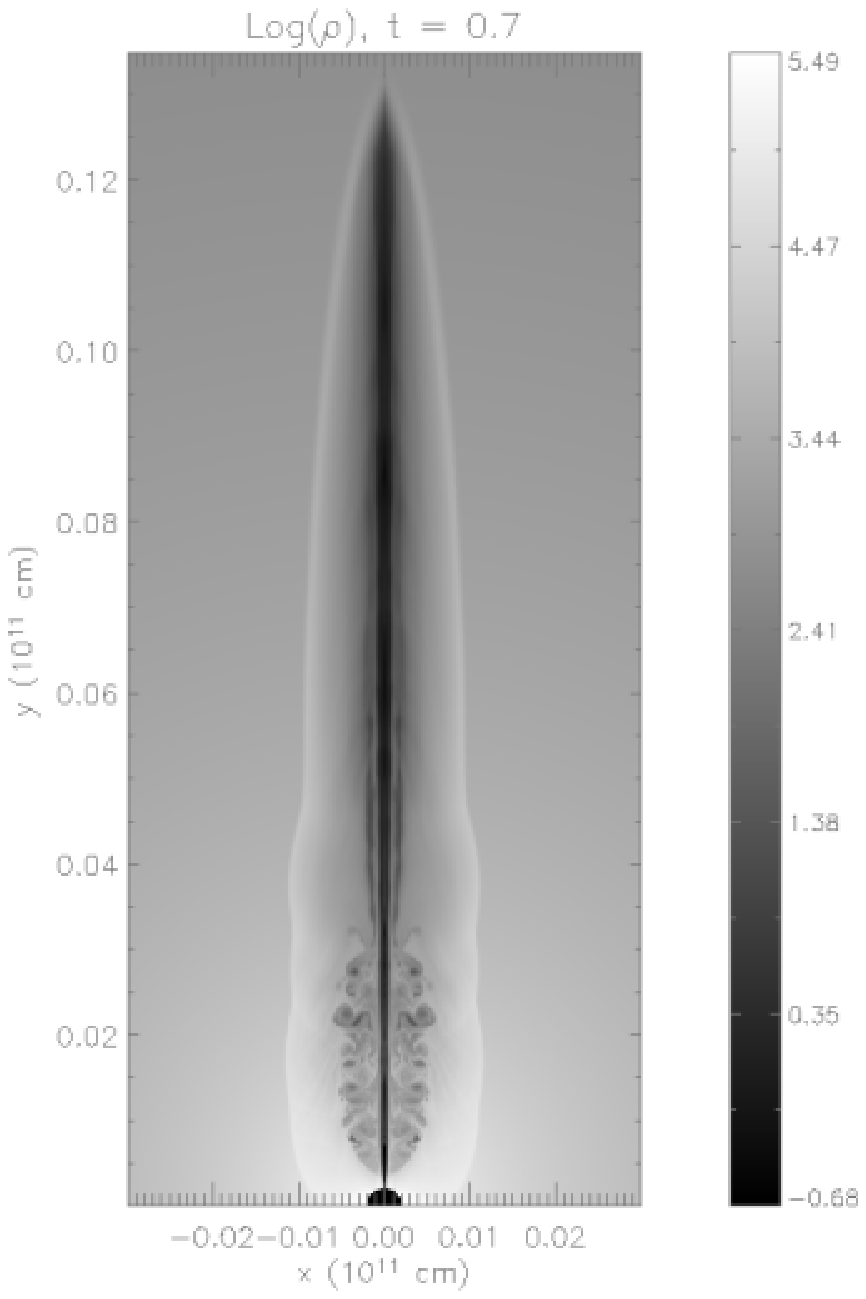}{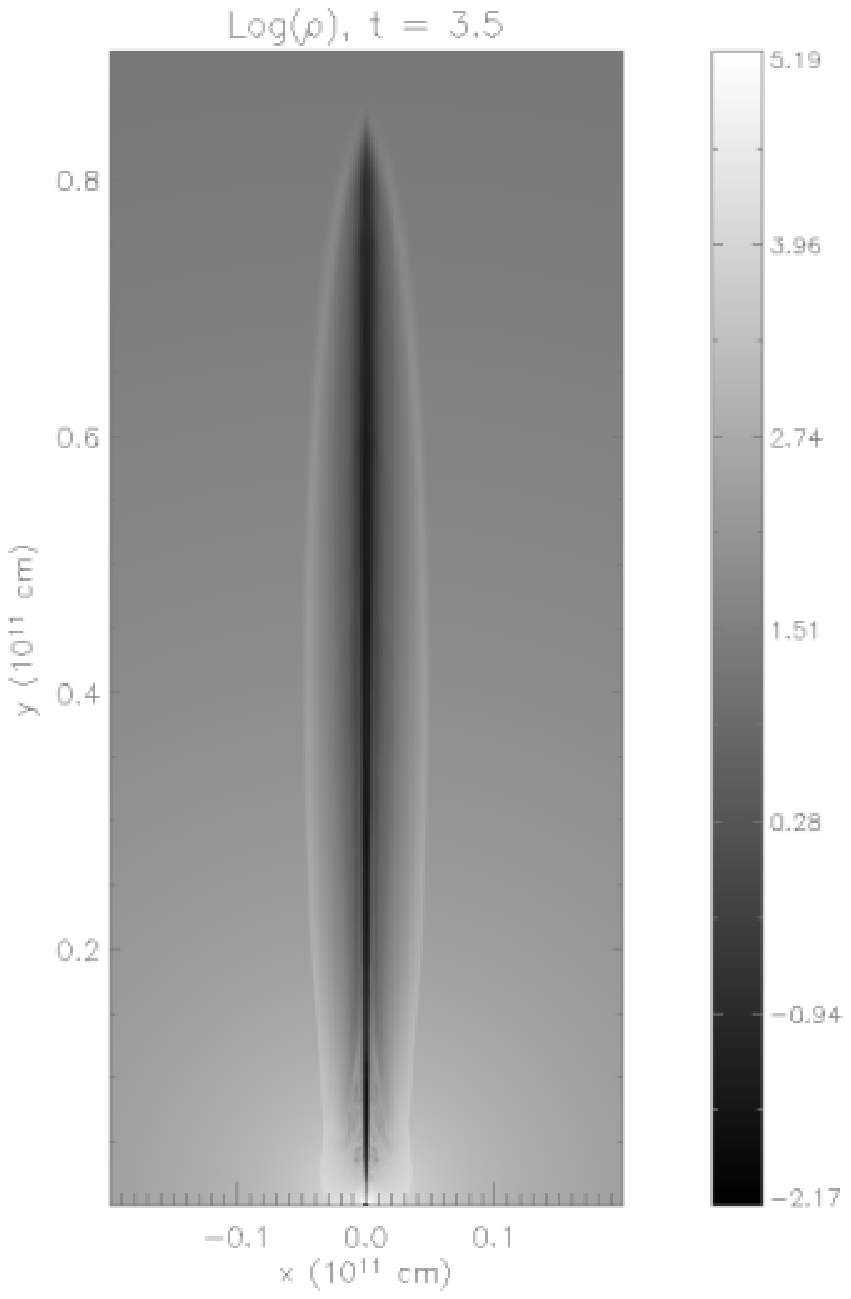}
 \caption{\footnotesize Density distribution at $t=0.7$ s (left panel)
          and $t=3.5$ s (right panel) for the gamma ray burst
          problem. The grid is unevenly spaced with a higher resolution close to the polar axis.
          The low density channel along the polar axis is evident at
          both times. The result can be compared to \cite{ZWmF03}, Figure 2.\label{fig:grb}}
\end{figure}

The results of our simulation are shown in
Figure \ref{fig:grb}. During the early stages ($t\lesssim 1$ s, left
panel in Fig. \ref{fig:grb}) the jet beam is quickly collimated by
the high pressure of the near stellar environment, $p_m/p_b \approx
263.5$. By $t=3.5$ s (right panel in Fig. \ref{fig:grb}), the jet has
cleared a low-density, high-velocity thin funnel along the polar axis
and remains narrowly collimated with a very thin featureless cocoon.
The latter property reflects the fact that ultra-relativistic jets are
less prone to Kelvin-Helmholtz instabilities \citep{FTZ78,MMFIM97}. 
This behavior is typical for supersonic light jets
characterized by low Mach numbers ($M_b \approx 1.82$ and
$\rho_b/\rho_m|_{r=r_0} \approx 1.85\cdot 10^{-5}$ for the present
case). One should also be aware that our simulation includes a certain
amount of numerical diffusion which limits the resolution of small scale
structures in our model. This is specially true at large radii where
the resolution of our mesh is coarser (the zone width increases
geometrically with radius) and the geometry is diverging (spatial resolution
decreases laterally).

Our results stay qualitative in agreement with the results of
\cite{ZWmF03} with only minor differences.  Specifically, the jet in
our model propagates about 50\% faster for $t < 0.7$ s but reaches
a similar distance by the final time ($t=3.5$ s). The morphological
differences between the two models are small, with the current model having
a slightly narrower beam and a less prominent cocoon.  These
discrepancies likely result from a combination of factors including
differences of the numerical methods, initial conditions, and physics
(we used simplified equation of state and neglect radiation effects).
Overall, however, the evolution of the jet in both models is very
similar.

\section{CONCLUSIONS}\label{sec:concl}
%
%
%
%
%

We presented a high-resolution numerical scheme for special
relativistic multidimensional hydrodynamics in general curvilinear
coordinates. A finite volume, Godunov-type formulation is used,
where volume averaged conserved quantities are evolved in time by
solving Riemann problems at each time step. The solver takes into
account non-vanishing tangential velocities at each zone interface and
assumes that the two non-linear waves are shocks (i.e. ``two-shock''
approximation is used). This greatly reduces the computational cost and
turns out to provide a reasonable approximation in the limit of weak
rarefaction waves (as long as the time integration is done explicitly
so that the time step is relatively small). The solution to the
Riemann problem is found iteratively and a new method of incorporating
a general EoS is presented.

We considered four different equations of state suitable for
relativistic hydrodynamics. A novel simple analytic formulation for
the relativistic perfect gas EoS has been presented.  Our new equation
of state recovers the exact solution \citep{Synge57} with accuracy
better than 4\%. This formulation is consistent with a special
relativistic formulation of the kinetic theory of gases and shows the
correct asymptotic behavior in the limit of very high and very low
temperatures. Since our EoS is given by a simple analytic expression,
the computational cost of the solution of the Riemann problem is
significantly reduced.

Multidimensional integration is done with the fully coupled
corner-transport upwind method \citep{Colella90,Saltzman94,MC02}.
Preserving symmetries of the problem is often difficult
especially when directionally split advection algorithm is used and
may require applying special procedures \citep{Aloy99}. Our choice of
an unsplit integrator makes the presented method free of such problems
and symmetries of the flow are perfectly preserved.

The calculation of the numerical fluxes requires solving 4 Riemann
problems in two dimensions and 12 Riemann problems in three dimensions
per cell per time step. This compares to 6 Riemann problems to be
solved for the unsplit second-order Runge-Kutta schemes (in three
dimensions) which in practice, however, usually require smaller time
steps. Second-order accuracy in time is achieved by using
characteristic projection operators.

Our implementation is verified in the case of strongly relativistic
flows with Lorentz factor in excess of $2\cdot 10^3$. Purely one-dimensional
test problems with non-vanishing tangential velocity are used to
demonstrate the formal correctness of the method. The implementation
is verified in two dimensions using a Riemann problem with two shocks
and two contact discontinuities. Performance of the new method is
illustrated using two astrophysical applications: a pressure-matched
light relativistic jet with Mach number $\approx 11$ and
$\Gamma\approx10$ in 2-D cylindrical geometry and the propagation of a
highly relativistic ($\Gamma > 50$) jet through the stellar atmosphere
in spherical geometry. The latter problem is motivated by studies of
relativistic jets in collapsars \citep{ZWmF03}.

\acknowledgements This work is supported in part by the
U.S. Department of Energy under Grant No.\ B523820 to the Center for
Astrophysical Thermonuclear Flashes at the University of Chicago.

\appendix 

\section{ORTHOGONAL CURVILINEAR COORDINATES}\label{app:geo_1}
%
%
%
%
%

In what follows we show how the geometrical source term $\A{S}_G$
(eq. [\ref{eq:cl_2b}]) can be calculated for an arbitrary orthogonal
system of coordinates.  We also provide explicit expressions for the
scale factors, face areas and zone volumes as required by the
conservative finite volume formulation presented in \S\ref{sec:algorithm}.
Expressions are given for the most commonly used
coordinate systems: cartesian, cylindrical and spherical.

Consider the rank-two tensor, $\A{T} = \sum_{a,b}T_{ab}\nv_a\nv_b$. The divergence
of $\A{T}$ is the vector
\begin{equation}
  \nabla\cdot\A{T} = \frac{1}{h_1h_2h_3}
  \sum_{d=1}^{3} \left(\pd{(h_2h_3T_{1d})}{x^1} +
                       \pd{(h_1h_3T_{2d})}{x^2} +
                       \pd{(h_1h_2T_{3d})}{x^3}\right)\nv_d + {\A{S}_G}(\A{T}) \,,
\end{equation}
where ${\A{S}_G}(\A{T})$ is the source term contributed by those versors
that do not have fixed orientation in space. The source term vector can be
expressed as
\begin{equation}
  {\A{S}_G}(\A{T}) = \sum_{d=1}^{3} \left(
 \frac{T_{1d}}{h_1}\pd{ \nv_d}{x^1} +
 \frac{T_{2d}}{h_2}\pd{ \nv_d}{x^2} +
 \frac{T_{3d}}{h_3}\pd{ \nv_d}{x^3} \right) \,,
\end{equation}
with components
\begin{equation}\label{eq:geo_src}
  \nv_d\cdot{\A{S}_G}(\A{T}) = \frac{1}{h_d} \sum_{a\ne d}
     \left(  \frac{T_{da}}{h_a}\pd{h_d}{x^a} - T_{aa}\pd{\log(h_a)}{x^d} \right)\,.
\end{equation}
The tensor $\A{T}$ appears as the dyad $\A{m}\A{v}$ in
equation (\ref{eq:cl_2b}).

First we consider Cartesian coordinates $(x^1,x^2,x^3) = (x,y,z)$. The
geometric scale factors are $h_x=h_y=h_z=1$ so that ${\A{S}_G}(\A{T})$
is identically zero. The cell volume is $\Delta\vol_{ijk} = \Delta
x_i\Delta y_j\Delta z_k$ and the surface areas of the six bounding
cell faces are
\begin{equation}
         A^x_{i\pm\HALF,j,k} = \Delta y_j\Delta z_k  \,, \quad
         A^y_{i,j\pm\HALF,k} = \Delta x_i\Delta z_k  \,, \quad
         A^z_{i,j,k\pm\HALF} = \Delta x_i\Delta y_j  \,.
\end{equation}

In cylindrical coordinates, $(x^1,x^2,x^3) \equiv
(r, \phi, z)$, the scale factors are
\begin{equation}
     h_r    =  1  \, , \quad
     h_\phi =  r  \, , \quad 
     h_z    =  1  \,.
\end{equation}
while the cell volume is
\begin{equation}\label{eq:vol_cyl}
     \Delta\vol_{i,j,k} \equiv \int_{r_{i-\HALF}}^{r_{i+\HALF}}
                   \int_{\phi_{j-\HALF}}^{\phi_{j+\HALF}}
                    \int_{z_{k-\HALF}}^{z_{k+\HALF}}  r \,drdzd\phi= 
                \left(\frac{r^2_{i+\HALF} - r^2_{i-\HALF}}{2}\right)\Delta \phi_j\Delta z_k \,.
\end{equation}
The areas of cell faces are
\begin{equation}
    A^r_{i\pm\HALF,j,k}    =  r_{i\pm\HALF} \Delta \phi_j\Delta z_k   \,, \quad
    A^\phi_{i,j\pm\HALF,k} =  \Delta r_i\Delta z_k                    \,, \quad
    A^z_{i,j,k\pm\HALF}    =  \left(\frac{r^2_{i+\HALF} - r^2_{i-\HALF}}{2}\right)\Delta \phi_j \,,
\end{equation}
and the geometrical source vector is
\begin{equation}
    {\A{S}_G}(\A{m}\A{v}) =  - \frac{m_\phi v_\phi}{r}  \nv_r
                                   + \frac{m_\phi v_r}{r}       \nv_\phi \,.
\end{equation}

In spherical coordinates, $(x^1,x^2,x^3) \equiv (r,\theta,\phi)$, the scale
factors are
\begin{equation}
    h_r      = 1  \,, \quad
    h_\theta = r  \,, \quad  
    h_\phi   = r\sin\theta \,,
\end{equation}
the cell volume is
\begin{equation}\label{eq:vol_sphe}
  \Delta\vol_{i,j,k} \equiv \int_{r_{i-\HALF}}^{r_{i+\HALF}}
                               \int_{\theta_{j-\HALF}}^{\theta_{j+\HALF}}
                               \int_{\phi_{k-\HALF}}^{\phi_{k+\HALF}}
                               r^2\sin\theta \,drd\theta d\phi 
                 =
    \left(\frac{r_{i+\HALF}^3 - r_{i-\HALF}^3}{3}\right)
    \left(\cos\theta_{j-\HALF} - \cos\theta_{j+\HALF}\right) \Delta\phi_k\,,
\end{equation}
and surface areas are
\begin{equation}
   A^r_{i\pm\HALF,j,k} = r^2_{i\pm\HALF} 
       \left(\cos\theta_{j-\HALF} - \cos\theta_{j+\HALF}\right) \Delta\phi_k \,,
\end{equation}
\begin{equation}
  A^\theta_{i,j\pm\HALF,k} = \left(\frac{r^2_{i+\HALF} - r^2_{i-\HALF}}{2}\right)
             \sin\theta_{j\pm\HALF} \Delta \phi_k   \,,
\end{equation}
\begin{equation}
  A^\phi_{i,j,k\pm\HALF}   = \left(\frac{r^2_{i+\HALF} - r^2_{i-\HALF}}{2}\right)\Delta\theta_j \,.
\end{equation}
In this case the geometrical source term is
\begin{equation}
    {\A{S}_G}(\A{m}\A{v}) = \left(-\frac{m_\theta v_\theta + m_\phi v_\phi}{r}\right)\nv_r
                           +\left(  \frac{m_\theta v_r - \cot\theta m_\phi v_\phi}{r}\right)\nv_\theta
                          + \left(  \frac{m_\phi   v_r + \cot\theta m_\phi v_\theta}{r}\right)\nv_\phi \,.
\end{equation}

\section{RECONSTRUCTION PROCEDURE} \label{app:PPM_curv}
%
%
%
%
%

In our parabolic reconstruction procedure we closely follow the
prescription of CW84 with only minor
modifications. The contact steepening algorithm, specific to PPM, has been
used only for one-dimensional tests and will not be reported here \citep[for
details see][]{MM96}. We also note that, as it was
pointed out by CW84, in order to preserve accuracy of the
scheme the interpolation should be formally done using the
conservative variables. Reconstruction based on primitive variables
offers, however, some advantages (preserves pressure positivity and in
case of relativistic hydrodynamics allows to satisfy kinematical and
thermodynamic constraints\footnote{Specifically, by requiring that  
$\Gamma < 2$, $|v| < 1$ and $h > 1$ one can show that 
$E > |m|$, and $D/E < \sqrt{1 - \A{m}^2/E^2}$
must be satisfied at all times.}) and is our method of choice.

Let $\bar{q}$ be the volume-average of some quantity $q$, 
$Q(\xi) = \int^\xi q(\xi')\, d\xi'$ its integral, where $\xi$ is a generic volume
coordinate. Here we adopt the implied notation $q \leftrightarrow
q_{i,j,k}$ whenever $i$, $j$ or $k$ are not present (the same
convention is adopted also for other three-dimensional quantities).
The reconstruction process begins with interpolating a quartic polynomial
through $Q_{i\pm2+\HALF}$,$Q_{i\pm1+\HALF}$,$Q_{i+\HALF}$.  This
polynomial is used to calculate the single-valued estimate
\begin{equation}
  q_{i+\HALF} = \left.\frac{dQ(\xi)}{d\xi}\right|_{\xi=\xi_{i+\HALF}}\,,
\end{equation}
at the zone interfaces. The explicit result of this construction is (CW84)
\begin{equation}\label{eq:ppm_LR}
   q_{i+\HALF} =  \bar{q}_i   + a_i\left(\bar{q}_{i+1} - \bar{q}_i\right)  
                        + b_i \overline {\delta q}_i
                        - c_i \overline {\delta q}_{i+1}\,,
\end{equation}
with
\begin{equation}
a_i = \frac{\Delta\xi_i}{\Delta\xi_i + \Delta\xi_{i+1}} +
      \frac{2(\Delta\xi_{i+1}c_i - \Delta\xi_{i}b_i)}
           {\Delta\xi_i + \Delta\xi_{i+1}} \,,
\end{equation}
\begin{equation}
b_i = \left(\frac{\Delta\xi_{i+1}}{\sum_{k=-1}^{k=2}\Delta\xi_{i+k}}\right)
      \frac{\Delta\xi_{i+1} + \Delta\xi_{i+2}}{\Delta\xi_i + 2\Delta\xi_{i+1}}
   \,, \quad
c_i = \left(\frac{\Delta\xi_i}{\sum_{k=-1}^{k=2}\Delta\xi_{i+k}}\right)
            \frac{\Delta\xi_{i-1} + \Delta\xi_i}{2\Delta\xi_i + \Delta\xi_{i+1}}\,.
\end{equation}

Monotonicity is enforced by limiting $q_{i+\HALF}$ to lie between
$\bar{q}_i$ and $\bar{q}_{i+1}$. This is achieved by using the limiter
\citep{VanLeer97}
\begin{equation}
   \overline{\delta q}_i = \left\{\begin{array}{ll} 
   \min(|\delta q_i|, 2|\bar{q}_i - \bar{q}_{i-1}|, 2|\bar{q}_i - \bar{q}_{i+1}|) \sgn(\delta q_i)
                   &  \textrm{if} \; (\bar{q}_{i+1} - \bar{q}_{i})(\bar{q}_{i} - \bar{q}_{i-1}) > 0 \,,
             \\ \noalign{\medskip}
    0        &  \textrm{otherwise}\,,
   \end{array}\right.
\end{equation}
where
\begin{equation}
  \delta q_i = 
  \frac{\Delta\xi_i}{\Delta\xi_{i-1} + \Delta\xi_i + \Delta\xi_{i+1}}
   \left[\frac{2\Delta\xi_{i-1} + \Delta\xi_i}{\Delta\xi_{i+1} + \Delta\xi_i} 
                 \left(\bar{q}_{i+1} - \bar{q}_i\right) 
  +    \frac{2\Delta\xi_{i+1} + \Delta\xi_i}{\Delta\xi_{i-1} + \Delta\xi_i} 
                 \left(\bar{q}_i - \bar{q}_{i-1}\right)\right] \,,
\end{equation}
is the average slope. At the end of this step we initialize
\begin{equation}
   q_{i+\HALF,L} = q_{i+\HALF,R} = q_{i+\HALF} \,,
\end{equation}
where $q_{i+\HALF,L}$ and $q_{i+\HALF,R}$ are 
the left and right limiting values of $q$ at the zone's right
interface.


Next, we constrain the interface values for zone $i,j,k$ to lie within
the extreme values found among all neighboring cells
\citep{Barth95}.
Construction for $q_{i+\HALF,L}$ (identical procedure
is followed to limit $q_{i-\HALF,R}$) proceeds as follows.  Let
$\hat{q}^{\max}$ and $\hat{q}^{\min}$ be, respectively, the maximum
and minimum value of $\bar{q}_{I,J,K}$ for $I = i-1, i, i+1$, $J =
j-1, j, j+1$, $K = k-1, k, k+1$. Then
\begin{equation}\label{eq:multid_lim}
  q_{i+\HALF,L} = \max\left(\hat{q}^{\min}, 
                  \min\left(\hat{q}^{\max}, q_{i+\HALF,L}\right)\right)\,.
\end{equation}
Notice that in two and three dimensions there is a total of 9 and 27
zones involved in the limiting step.

In the next step we limit the parabolic distribution of $q$ to ensure
that its profile remains monotonic in each cell. A first case when
monotonicity can be violated is when $\bar{q}$ is a local maximum or
minimum. In this case we revert to first order interpolation:
\begin{equation}
    q_{i+\HALF,L} \;\rightarrow\; \bar{q} 
         \qquad \textrm{if}\quad 
   (q_{i-\HALF,R} - \bar{q})(\bar{q} - q_{i+\HALF,L}) \le 0\,.
\end{equation}

Monotonicity can also be violated if the parabolic distribution
(\ref{eq:parabola}) has an extremum inside the zone. In this case one
of the interface values is adjusted so that the parabolic profile has
an extremum at the other interface. The modified distribution also has
to preserve the correct zone-average value. The final result is:
\begin{equation}
   q_{i-\HALF,R} \;\rightarrow\; 3\bar{q} - 2q_{i+\HALF,L} \qquad\textrm{if}\quad
   \left(q_{i+\HALF,L} - q_{i-\HALF,R}\right) 
   \left(\bar{q} - \frac{q_{i+\HALF,L} + q_{i-\HALF,R}}{2}\right) > 
   \frac{\left(q_{i+\HALF,L} - q_{i-\HALF,R}\right)^2}{6} \,,
\end{equation}
\begin{equation}
   q_{i+\HALF,L} \;\rightarrow\; 3\bar{q} - 2q_{i-\HALF,R} \qquad\textrm{if}\quad
   \left(q_{i+\HALF,L} - q_{i-\HALF,R}\right) 
   \left(\bar{q} - \frac{q_{i+\HALF,L} + q_{i-\HALF,R}}{2}\right) <
   -\frac{\left(q_{i+\HALF,L} - q_{i-\HALF,R}\right)^2}{6} \,.
\end{equation}

It should also be mentioned that the reconstruction algorithm may 
occasionally fail to respect the condition $v_1^2 + v_2^2 + v_3^2 < 1$.  
To prevent this from happening, interpolation of the 
velocity components reverts to first order whenever  
the total velocity exceeds the bounds provided by the neighboring cells, 
in a way similar to equation (\ref{eq:multid_lim}). 

\subsection{Dissipation Algorithm}\label{app:dissip}
%
%
%
%
%

Interpolation profiles are modified in presence of strong shocks to
prevent unphysical oscillations. This is achieved by the flattening
algorithm in which the interface values are modified as
\begin{equation}
  q_{i+\HALF,L} = \chi q_{i+\HALF,L} + (1 - \chi)\bar{q}\,,
\end{equation}
\begin{equation}
  q_{i-\HALF,R} = \chi q_{i-\HALF,R} + (1 - \chi)\bar{q}\,,
\end{equation}
where $\chi$ is a multi-dimensional flattening parameter, $0\le\chi\le1$.
Note that for $\chi = 0$ the accuracy of the method reduces to first
order.
The flattening parameter is computed as \citep{MC01, MC02}
\begin{equation}
   \chi = \min_{I,J,K}\left(\tilde{\chi}_{1,I,j,k}, \tilde{\chi}_{2,i,J,k}, 
                    \tilde{\chi}_{3,i,j,K}\right) \,,
\end{equation}
where the minimum is taken over all $I,J,K$ in the range $i-1\le I\le
i+1$, $j-1\le J\le j+1$, $k-1\le K\le k+1$.  The coefficients
$\tilde{\chi}_1$, $\tilde{\chi}_2$ and $\tilde{\chi}_3$ are
one-dimensional. In what follows we describe the procedure only for
$\tilde{\chi}_1$; the remaining two values are calculated in the same
way.
First, we introduce a measure of the shock width
\begin{equation}
  \beta_1 = \frac{|p_{i+1} - p_{i+1}|}{|p_{i+2} - p_{i+2}|} \,,
\end{equation}
and shock strength
\begin{equation}
  Z_1 = \frac{|p_{i+1} - p_{i-1}|}{\min(p_{i-1}, p_{i+1})} \,.
\end{equation}
Next we define
\begin{equation}
  \tilde{\chi}^{min}_1 = \max\left(0, \min
    \left(1, \frac{\beta^{max} - \beta_1}{\beta^{max} - \beta^{min}}\right)\right) \,,
\end{equation}
so that flattening is not applied for $\beta < \beta^{min}$.  The
one-dimensional flattening parameter is then restricted only to the
regions where the flow undergoes compression and its value depends on the
shock strength:
\begin{equation}
  \tilde{\chi}_1 = \left\{\begin{array}{ll}
 \DS \max\left(\tilde{\chi}^{\min}_1, \min
         \left(1, \frac{Z^{max} - Z_1}{Z^{max}- Z^{min}}\right)\right) 
   & \qquad \textrm{if} \quad v_{1,i+1} < v_{1,i}     \,,  \\ \noalign{\medskip}
  1  & \qquad \textrm{otherwise}                      \,.  \\
  \end{array}\right.
\end{equation}
For the present study we follow recommendation of
\cite{MC02} and adopt $\beta^{max} = 0.85$, $\beta^{min}
= 0.75$, $Z^{max} = 0.75$, $Z^{min} = 0.25$.

Slope flattening is combined with the introduction
of an explicit diffusive flux, in order to further 
reduce spurious oscillations behind strong shocks.
To this purpose, the numerical fluxes (eq. [\ref{eq:unsplit}]) in the 
final conservative update are augmented as
\begin{equation}
  \A{F}_{i+\HALF} \;\rightarrow\; \A{F}_{i+\HALF} + k_\nu \left(\A{U} - \A{U}_{i+1}\right) \,.
\end{equation}
Here
\begin{equation}
  k_\nu = \alpha\max\left(-\mathcal{D}_{i+\HALF},0\right) \,,
\end{equation}
where $\alpha$ is typically set to $0.1$. 
$\mathcal{D}_{i+\HALF}$ is a measure of the convergence of the flow at 
the zone interface $i+\HALF$. In cartesian coordinates with a uniform 
zone spacing ($\Delta x=\Delta y= \Delta z$), 
$\mathcal{D}_{i+\HALF}$ is a discrete undivided difference approximation to the 
multidimensional divergence of $\A{v}$. 
In order to compute $\mathcal{D}_{i+\HALF}$, we define a
measure of the convergence of the flow at the cell corners as
\begin{equation}\begin{array}{cl}
  \mathcal{D}_{i+\HALF,j+\HALF,k+\HALF} =  
 \DS  \frac{1}{4}&\DS \left[ \sum_{J=j,j+1\atop K=k,k+1} \left( v_{1,i+1,J,K} - v_{1,i,J,K}\right)\right.
 \DS + \left.\sum_{I=i,i+1\atop K=k,k+1} \left( v_{2,I,j+1,K} - v_{2,I,j,K}\right) \right.\\
                              & 
 \DS +\left.\sum_{I=i,i+1\atop J=j,j+1} \left( v_{3,I,J,k+1} - v_{3,I,I,k}\right)\right] \,,
\end{array}\end{equation}
so that $\mathcal{D}_{i+\HALF}$ is obtained by a simple 
average procedure
\begin{equation}
  \mathcal{D}_{i+\HALF} = \frac{1}{4}\left(
       \mathcal{D}_{i+\HALF,j+\HALF,k+\HALF}
     + \mathcal{D}_{i+\HALF,j-\HALF,k+\HALF}
     + \mathcal{D}_{i+\HALF,j+\HALF,k-\HALF} 
     + \mathcal{D}_{i+\HALF,j-\HALF,k-\HALF}\right) \,.
\end{equation}

%
%

%
%

\clearpage

\clearpage

\clearpage

\begin{figure}
 \epsscale{0.6}
 \plotone{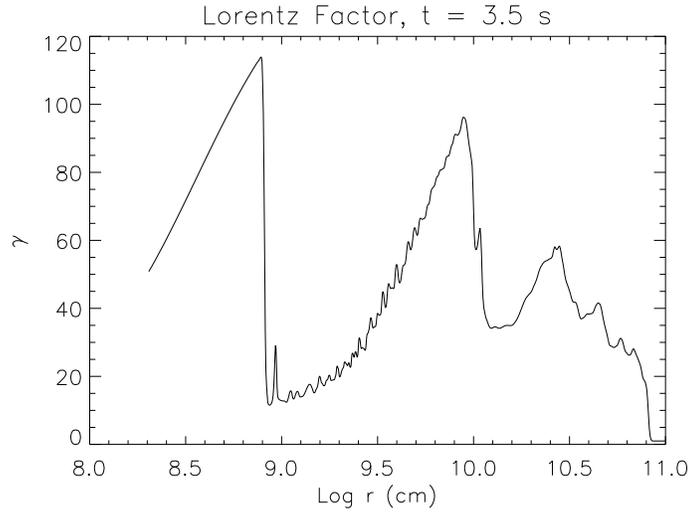}
 \caption{\footnotesize Lorentz factor for the gamma ray burst 
           problem at $t=3.5$ s. The reverse shock is located at 
           $\log r \approx 8.9$. \label{fig:grb_lorentz}}
\end{figure}

%
%
%

\clearpage
\begin{deluxetable}{cccc}
\tablecaption{Expressions for the different EoS, square of sound speed and solutions to the
              Taub adiabat  for the four different cases presented in the text; the coefficients
              $a$, $b$ and $c$ for the $ID$ and $TM$ EoS are given by equations
              (\ref{eq:abc_ID}) and (\ref{eq:abc_TM}).
              The $*$ case has to be solved numerically.\label{tab:EoS1}}
\tablewidth{0pt}
\tablehead{
\colhead{EoS} & \colhead{$h(\Theta)$} & \colhead{$c_s^2$} & \colhead{Taub adiabat}} 
\startdata
ID & $\DS 1 + \Gamma_r\Theta$     & 
     $\DS \frac{\Gamma\Theta}{h}$ &
     $\DS h =\frac{ -b + \sqrt{b^2 - 4ac}}{2a}$ \\ \noalign{\bigskip}
RP & $\DS \frac{K_3(1/\Theta)}{K_2(1/\Theta)}$                                       & 
     $\DS \frac{\Theta}{h} \frac{1 + 5h\Theta - h^2}{1 + 5h\Theta - h^2 - \Theta^2}$ & *\\ \noalign{\bigskip}
IP & $\DS 2\Theta + \sqrt{4\Theta^2 +1}$    &
     $\DS \frac{2\Theta}{h+2\Theta}$        & 
     $\DS h^2 = h_S^2 + \frac{4h_S\tau_S[p^2]}{3p + p_S}$ \\ \noalign{\bigskip}
TM & $\DS \frac{5}{2}\Theta + \sqrt{\frac{9}{4}\Theta^2 +1}$ &
     $\DS \frac{\Theta}{3h}\frac{5h-8\Theta} {h-\Theta}$     &
     $\DS [h\tau] =  \frac{2c}{|b| + \sqrt{b^2 - 4ac}}$  \\
\enddata

\end{deluxetable}

\begin{deluxetable}{ccc}
\tablecaption{Expressions for the mass flux $j$, and derivative $d(h\tau)/dp$
              for the four  different EoS presented in the text; here $w \equiv h\tau$.
              Notice that for all but the RP case, the expressions for $j^2$ is numerically well
              behaved for $[p] \to 0$. \label{tab:EoS2}}
\tablewidth{0pt}
\tablehead{
\colhead{EoS} & \colhead{$j^2$} & \colhead{$d(h\tau)/dp$} }
\startdata
ID & $\DS  \frac{\Gamma_r p}{\Gamma_r w_S +(w + w_S)\left(\frac{1}{h+h_S} - 1\right)}$ & 
     $\DS  \frac{w + w_S  - \frac{2h}{2h-1} \Gamma_r w}{\frac{2h}{2h - 1}\Gamma_rp - [p]}$ \\ \noalign{\bigskip}
RP & $\DS -\frac{pp_S[p]}{p_S[h\Theta] - h_S\Theta_S[p]}$   &
     $\DS  \frac{(w + w_S)(h - \dot{h}\Theta - 2h\dot{h}w)}{ 2\dot{h}hp - [p](h - \dot{h}\Theta)}$\\ \noalign{\bigskip}
IP & $\DS  \frac{4p}{3w_S - w}$   &
     $\DS  \frac{w_S - 3w} {3p + p_S}$ \\  \noalign{\bigskip}
TM & $\DS  \frac{|b| + \sqrt{b^2 - 4ac}}{2w_S(h^2_S  +  2w_S p + 2)}$ &
     $\DS  \frac{w(3h^2 - 3h\Theta + 1) - w_S(2h^2 - 1 - 5h\Theta)}{1 + [p](1 + 5h\Theta - 2h^2)}$\\
\enddata

\end{deluxetable}

\begin{deluxetable}{cccc} 
\tablecolumns{4} 
\tablewidth{0pc} 
\tablecaption{Discrete L-1 norms of the conserved quantities
              for the shock tube problem (\emph{P1}).
              \label{tab:L-1error}}
\tablehead{\colhead{$\Delta x^{-1}$} & \colhead{$\|\epsilon_D\|_1$} & \colhead{$\|\epsilon_m\|_1$}  & \colhead{$\|\epsilon_E\|_1$}}
\startdata
   50  &  $0.1620$  &  $0.2160$  &  $0.1560$  \\
  100  &  $0.0920$  &  $0.1420$  &  $0.1000$  \\
  200  &  $0.0505$  &  $0.0792$  &  $0.0558$  \\
  400  &  $0.0298$  &  $0.0436$  &  $0.0319$  \\
  800  &  $0.0118$  &  $0.0264$  &  $0.0206$  \\
 1600  &  $0.0115$  &  $0.0149$  &  $0.0126$  \\
\enddata
\tablecomments{\footnotesize  Errors are computed according to $\|\epsilon_q\|_1 = \sum_i \Delta x |q_i -
                              Q(x_i)|$ (as in \cite{MM96}), where $q$ is one of $D$, $m$, $E$,
                              and $Q(x_i)$ is the exact solution at $x = x_i$.}
\end{deluxetable}

\begin{deluxetable}{ccccc} 
\tablecolumns{4} 
\tablewidth{0pc}
\tablecaption{Relative global errors \citep[computed as in][]{Aloy99} for the Relativistic Spherical Shock 
              Reflection (RSSR) test problem for different inflow velocities at $t = 2$. 
              Computations have been performed on a $81^3$ grid.
              \label{tab:rssr}}
\tablehead{\colhead{$\nu$} & \colhead{$\gamma$} & \colhead{$\|\epsilon_\rho\|_1$} & \colhead{$\|\epsilon_v\|_1$}  & \colhead{$\|\epsilon_p\|_1$}}
\startdata
   $10^{-1}$  & $2.29$   &   $0.18$  &  $0.01$  &  $0.29$  \\
   $10^{-3}$  & $22.37$  &   $0.20$  &  $0.03$  &  $0.17$  \\
   $10^{-5}$  & $223.6$  &   $0.26$  &  $0.04$  &  $0.20$  \\
   $10^{-7}$  & $2236.1$ &   $0.21$  &  $0.04$  &  $0.24$  \\
\enddata
\tablecomments{\footnotesize The inflow velocity is given by $v_{in} = \nu - 1$. The corresponding 
                             Lorentz factor is given in the second column. The first two
                             cases have been run with CFL = $0.4$, while CFL = $0.1$ has been used
                             for the last two cases.}
\end{deluxetable}


\begin{thebibliography}{}

\bibitem[Ackermann et al.(2001)]{Ackermann+01}
  Ackermann, K. H., et al.\
  2001, PRL, 86, 402

\bibitem[Aloy et al.(1999)]{Aloy99}
  Aloy, M.~A., Ib{\' a}{\~ n}ez, J.~M. \& Mart{\'{\i}}, J.~M.\ 
  1999, \apjs, 122, 151 

\bibitem[Aloy et al.(2000)]{Aloy00}
  Aloy, M.~A., M{\" u}ller, E., Ib{\' a}{\~ n}ez, J.~M., Mart{\'{\i}}, J.~M. \& MacFadyen, A.\ 
  2000, \apjl, 531, L119 

\bibitem[Aloy et al.(2003)]{Aloy+03}
  Aloy, M.~A., et al.\ 
  2003, \apjl, 585, L109 

\bibitem[Anile (1989)]{Anile89}
  Anile, A.~M.\ 
  1989, Relativistic Fluids and Magneto-fluids (Cambridge: Cambridge University Press), 55

\bibitem[Balsara(1994)]{Balsara94}
  Balsara, D.~S.\ 
  1994, J. Comput. Phys., 114, 284

\bibitem[Barth(1995)]{Barth95}
  Barth, T. J. 
  1995, Aspects of Unstructured Grids and Finite-Volume Solvers for Euler and Navier-Stokes Equations, 
  VKI/NASA/AGARD Special Course on Unstructured Grid Methods for Advection Dominated Flows (AGARD Publ. R-787) 
  (Belgium: Von Karmen Inst. for Fluid Dynamics)

\bibitem[Begelman et al.(1984)]{Begelman+84}
  Begelman, M.~C., Blandford, R.~D. \& Rees, M.~J.\
  1984, Rev. Mod. Phys., 56, 255

\bibitem[Blondin \& Lufkin(1993)]{BL93}
  Blondin, J.~M.~\& Lufkin, E.~A.\ 
  1993, \apjs, 88, 589 

\bibitem[Bogovalov et al.(2005)]{Bogovalov+05}
  Bogovalov, S.~V., et al.\
  2005, \mnras, 358, 705

\bibitem[Calder et al.(2002)]{Calder_etal02}
  Calder, A.~C., et al.\ 
  2002, \apjs, 143, 201 

\bibitem[Colella(1982)]{Colella82}
  Colella, P.\ 
  1982, SIAM J. Sci. Stat. Comput., 3, 76

\bibitem[Colella \& Woodward(1984)]{CW84}
  Colella, P.~\& Woodward, P.~R.\ 
  1984, J. Comput. Phys., 54, 174

\bibitem[Colella(1985)]{Colella85}
  Colella, P.\ 
  1985, SIAM J. Sci. Stat. Comput., 6, 104

\bibitem[Colella \& Glaz(1985)]{CG95}
  Colella, P.~\& Glaz, H.M.  
  1985, J. Comput. Phys., 59, 264

\bibitem[Colella(1990)]{Colella90}
  Colella, P. 
  1990, J. Comput. Phys., 87, 171

\bibitem[Courant et al.(1928)]{CFL28}
  Courant, R., Friedrichs, K.~O. \& Lewy, H.\
  1928,  Math. Ann., 100, 32

\bibitem[Dai \& Woodward(1997)]{DW97}
  Dai, W.~\& Woodward, P.~R.\
  1997, SIAM J. Sci. Comput., 18, 982

\bibitem[Del Zanna \& Bucciantini(2002)]{dZB02}
  Del~Zanna, L.~\& Bucciantini, N.\ 
  2002, \aap, 390, 1177

\bibitem[Del Zanna et al.(2004)]{DelZanna+04}
  Del~Zanna, L., Amato, E. \& Bucciantini, N.\
  2004, \aap, 421, 1063

\bibitem[Dolezal \& Wong(1995)]{Dol_Wong95}
  Dolezal, A.~\& Wong, S.~M.\
  1995, J. Comput. Phys., 120, 266

\bibitem[Donat et al.(1998)]{DFIM98}
  Donat, R., Font, J.~A., Ib{\' a}{\~ n}ez, J.~M. \& Marquina, A.\ 
  1998, J. Comput. Phys., 146, 58

\bibitem[Falle \& Komissarov(1996)]{FK96}
  Falle, S.~A.~E.~G~\& Komissarov, S.~S.\ 
  1996, \mnras, 278, 586

\bibitem[Ferrari et al.(1978)]{FTZ78}
  Ferrari, A., Trussoni, E. \& Zaninetti, L.\ 
  1978, \aap, 64, 43

\bibitem[Frail et al.(2001)]{Frail01}
  Frail, D.~A., et al.\ 
  2001, \apjl,  562, L55

\bibitem[Koide et al.(1999)]{Koide+99}
  Koide, S., Shibata, K. \& Sakai, J.\
  1999, \apj, ApJ, 522, 727

\bibitem[Lax \& Liu(1998)]{LL98}
  Lax, P.~D.~\& Liu, X.-D.\
  1998, SIAM J. Sci. Comput., 19, 319

\bibitem[Landau \& Lifshitz(1959)]{Lan_Lif59}
  Landau, L.~D.~\& Lifshitz, E.~M.\ 
  1959, Fluid Mechanics (New York: Pergamon)

\bibitem[Marquina et al.(1992)]{MMIMD92}
  Marquina, A., Mart{\'{\i}}, J.M., Ib{\' a}{\~ n}ez, J.~M.,
  Miralles, J.~A. \& Donat, R.\ 
  1992, \aap, 258, 566

\bibitem[Mart{\'{\i}} \& M{\" u}ller(1994)]{MM94}
  Mart{\'{\i}}, J.~M.~\& M{\" u}ller, E.\
  1994, J. Fluid Mech., 258, 317

\bibitem[Mart{\'{\i}} \& M{\" u}ller(1996)]{MM96}
  Mart{\'{\i}}, J.~M.~\& M{\" u}ller, E.\
  1996, J. Comput. Phys., 123, 1

\bibitem[Mart{\'{\i}} et al.(1997)]{MMFIM97}
  Mart{\'{\i}}, J.~M., M{\" u}ller, E., Font, J.~A.,
  Ibanez, J.~M.~A. \& Marquina, A.\ 
  1997, \apj, 479, 151 

\bibitem[Mart{\'{\i}} \& M{\" u}ller(2003)]{MM03}
  Mart{\'{\i}}, J.~M.~\& M{\" u}ller, E.\ 
  2003, Living Reviews in Relativity, 6, 7 

\bibitem[Meier et al.(2001)]{Meier+01}
  Meier, D. L., Koide, S. \& Uchida, Y.\
  2001, Science, 291, 84

\bibitem[Miller \& Colella(2001)]{MC01}
  Miller, G.~H.~\& Colella, P.\ 
  2001, J. Comput. Phys., 167, 131

\bibitem[Miller \& Colella(2002)]{MC02}
  Miller, G.~H.~\& Colella, P.\
  2002, J. Comput. Phys., 183, 26

\bibitem[Mizuno et al.(2004)]{Mizuno+04}
  Mizuno, Y., et al.\
  2004, \apj, 615, 389

\bibitem[Molnar \& Huovinen(2004)]{Molnar+04}
  Molnar, D. \& Huovinen, P.\
  2004, Phys. Rev. Lett., 94, 012302

\bibitem[Morita et al.(2002)]{Morita+02}
  Morita, K., Muroya, S., Nonaka, C. \& Hirano, T.\
  2004, Phys. Rev. C, 66, 054904

\bibitem[Norman \& Winkler(1986)]{NW86}
  Norman, M. L. \& Winkler, K.-H.\
  1986, in Astrophysical Radiation Hydrodynamics, 
  ed. K.-H. Winkler \& M. Norman (Dordrecht: Kluwer), 449 

\bibitem[Pons et al.(2000)]{PMM00}
  Pons, J.A., Mart{\'{\i}}, J.M. \& M{\" u}ller, E.\
  2000, J. Fluid Mech., 42, 125

\bibitem[Rezzolla et al.(2003)]{RZP03}
  Rezzolla, L., Zanotti, O., Pons, J.A.,\
  2003, J. Fluid Mech., 479, 199

\bibitem[Saltzman(1994)]{Saltzman94}
  Saltzman, J.\
  1994, J. Comput. Phys., 115, 153

\bibitem[Schulz et al.(1993)]{SCG93}
  Schulz-Rinne, C.~W., Collins, J.~P. \& Glaz, H.~M.\
  1993, SIAM J. Sci. Comp., 14, 1394 

\bibitem[Sokolov et al.(2001)]{SZS01}
  Sokolov, I.~V., Zhang, H.-M. \&  Sakai, J.~I.\ 
  2001, J. Comput. Phys., 172, 209 

\bibitem[Strang(1968)]{Strang68}
  Strang, G.,
  1968, SIAM J. Num. Anal., 5, 506

\bibitem[Synge(1957)]{Synge57}
  Synge, J.~L.\
  1957, The relativistic Gas, North-Holland Publishing Company

\bibitem[Taub(1948)]{Taub48}
  Taub, A.~H.\
  1948, Physical Review, 74, 328 
  
\bibitem[Toro(1997)]{Toro97}
  Toro, E.~F.\ 
  1997, Riemann Solvers and Numerical Methods for Fluid Dynamics,
  Springer-Verlag, Berlin

\bibitem[VanLeer(1997)]{VanLeer97}
  VanLeer, B.\ 
  1997, J. Comput. Phys., 135, 229 

\bibitem[Weinberg(1972)]{Weinberg72}
  Weinberg, S.\ 
  1972,  Gravitation and Cosmology, (New York: Wiley)

\bibitem[Wilson(1972)]{Wilson72}
  Wilson, J.~R.\ 
  1972, \aap, 173, 431

\bibitem[Woodward \& Colella(1984)]{WC84}
  Woodward, P.~R.~\& Colella, P.\ 
  1984, J. Comput. Phys., 54, 115

\bibitem[Woosley \& MacFayden(1999)]{WmF99}
  Woosley, S.~E.~\& MacFayden, A.~I.\ 
  1999, \aap, 138, 499 

\bibitem[Zhang et al.(2003)]{ZWmF03}
  Zhang, W., Woosley, S.~E. \& MacFayden, A.~I.\ 
  2003, \apj, 586, 356 

\end{thebibliography}
\end{document}